\documentclass[%
 preprint,
 superscriptaddress,
 amsmath,amssymb,
 aps,
 showkeys,
 titlepage
]{revtex4-2}

\usepackage{graphicx}
\usepackage{dcolumn}
\usepackage{bm}
\usepackage[colorlinks = true,
            linkcolor = blue,
            urlcolor  = blue,
            citecolor = red,
            anchorcolor = blue]{hyperref}

\begin{document}


\title{Indirect Exchange Interaction Leads to Large Lattice Contribution to Magnetocaloric Entropy Change} 

\author{Lokanath Patra}
\affiliation{Department of Mechanical Engineering, University of California, Santa Barbara, CA 93106, USA}

\author{Bolin Liao}
\email{bliao@ucsb.edu} \affiliation{Department of Mechanical Engineering, University of California, Santa Barbara, CA 93106, USA}

\date{\today}

\begin{abstract}
Materials with a large magnetocaloric response are highly desirable for magnetic cooling applications. It is suggested that a strong spin-lattice coupling tends to generate a large magnetocaloric effect, but no microscopic mechanism has been proposed. In this work, we use spin-lattice dynamics simulation to examine the lattice contribution to the magnetocaloric entropy change in bcc iron (Fe) and hcp gadolinium (Gd) with exchange interaction parameters determined from \textit{ab-initio} calculations. We find that indirect Ruderman–Kittel–Kasuya–Yosida (RKKY) exchange interaction in hcp Gd leads to longer-range spin-lattice coupling and more strongly influences the low-frequency long-wavelength phonons. This results in a higher lattice contribution towards the total magnetocaloric entropy change as compared to bcc Fe with short-range direct exchange interactions. Our analysis provides a framework for understanding the magnetocaloric effect in magnetic materials with strong spin-lattice couplings. Our finding suggests that long-range indirect RKKY-type exchange gives rise to a larger lattice contribution to the magnetocaloric entropy change and is, thus, beneficial for magnetocaloric materials.
\end{abstract}

                            
\maketitle



\section{Introduction}

Magnetic refrigeration is based on the magnetocaloric effect (MCE), which is the material's ability to heat (cool) when magnetized (demagnetized) in an adiabatic process.~\cite{pecharsky1999magnetocaloric,balli2017advanced, shen2009recent, franco2018magnetocaloric} The MCE originates from the magnetic order-disorder transition induced by an external magnetic field and the associated entropy change. Understanding and designing materials with a strong MCE are of both great scientific and technological importance. Fundamentally, MCE provides a convenient probe to examine the interplay between magnetism and other excitations in condensed matter systems.~\cite{balz2019finite,czajka2021oscillations} Technologically, MCE has been widely adopted to obtain cryogenic temperatures in space missions,~\cite{shirron2012design} observatory astronomy~\cite{britt1981adiabatic} and scientific experimentation,~\cite{sato2016tiny} where compact and reliable cooling solutions are required. Magnetic refrigeration near room temperature has also been considered as an environmentally friendly alternative to conventional refrigeration based on vapor compression cycles.~\cite{aprea2015magnetic}   MCE materials can be characterized by the isothermal entropy change $\Delta$S, which measures the change in the equilibrium entropy of a material as a result of an externally applied magnetic field. Under isothermal conditions, the entropy change $\Delta$S manifests itself as the amount of heat released or absorbed by the material when an external magnetic field is applied or removed. Therefore, $\Delta$S is a metric for the cooling capacity of an MCE material. Improvement in the overall performance of a magnetic cooling system is primarily dependent on the isothermal entropy change of the magnetic refrigerant material and is of significant current scientific interest, particularly since the discovery of the giant MCE in Gd$_5$(Si$_2$Ge$_2$) in 1997 by Pecharsky and Gschneidner.~\cite{pecharsky1997giant}

Magnetocaloric materials with a strong coupling between spin and lattice degrees of freedom are known to exhibit a large MCE. Prominent examples are the first-order MCE materials, where the magnetic transition is associated with a first-order structural phase transition. Notable first-order MCE materials include Gd$_5$(Si$_x$Ge$_{4-x}$) series,~\cite{choe2003nanoscale, pecharsky1997giant, magen2003pressure}, MnAs$_{1-x}$Sb$_x$ alloys,~\cite{wada2001giant} La–Fe–Si-based alloys,~\cite{shao2022depth, bao2012magnetocaloric} Mn–Fe–P-based alloys~\cite{tegus2002transition} and Ni-Mn-based Heusler compounds.~\cite{krenke2005inverse, krenke2007magnetic} The observed large MCE is governed by the concurrent magnetic and structural phase transition as a result of the strong spin-lattice coupling phenomenon since the external applied magnetic field can simultaneously change the magnetic and lattice entropy in these materials. The lattice contribution has been reported to be as high as $50\%-60\%$ or more of the total entropy change in the materials undergoing a magnetostructural or magnetoelastic transition,~\cite{pecharsky2003massive,morellon2004pressure,wada2009pressure} and the magnitude of the lattice entropy change ($\Delta$S$_L$) is closely related to the volume change ($\Delta V/V$) during the phase transition, which can create mechanical issues and pose practical challenges in applying first-order MCE materials.~\cite{kuz2007factors} Even in conventional second-order MCE materials, a strong spin-lattice coupling is usually an indicator of a strong MCE. Based on this observation, Bocarsly et al. performed a computational screening of MCE materials using the spin-dependent lattice parameter as a proxy, which is an approximate computational measure of spin-lattice coupling.~\cite{bocarsly2017simple} Despite the abundance of empirical evidence, an atomic-level understanding of the relationship between spin-lattice coupling and the MCE is currently lacking. In particular, it is unclear what microscopic mechanisms are responsible for the strong spin-lattice coupling and the associated high MCE. In this light, atomistic computational methods to quantify the entropy contributions from the spin and the lattice degrees of freedom separately are of pivotal importance for the discovery and optimization of the MCE materials.  

The isothermal entropy change in magnetocaloric materials can be calculated from their magnetizations as functions of temperature and applied magnetic field, following the thermodynamic Maxwell relation~\cite{pecharsky1999magnetocaloric}:
\begin{equation}
\Delta S (T, \Delta H) = \mu_0 \int_{H_i}^{H_f} \left( \frac{\partial M(T,H)}{\partial T} \right)_H\ dH,
\label{eqn:entropy_change}
\end{equation}
where $\Delta H = H_f - H_i$ is the change of the applied external field ($H_f$ and $H_i$ are final and initial fields, respectively), $\mu_0$ is the Bohr magneton, $M$ is the magnetization and $T$ is the temperature.
The field- and temperature-dependent magnetization can be obtained by simulating the dynamics of atomic spins in magnetic materials using atomic spin dynamics (ASD) simulations with magnetic exchange interaction parameters ($J_{ij}$, where $i$ and $j$ label the interacting spins) calculated from \textit{ab initio} methods.~\cite{antropov1995ab, skubic2008method,ma2014dynamic}. However, these simulations ignore the effect of thermal fluctuations of the atoms (lattice vibrations) at finite temperatures and, thus, cannot capture the spin-lattice coupling effect and the resultant lattice contribution to the magnetocaloric entropy change ($\Delta S_L$). Hence, ASD simulations need to be modified to account for the lattice vibrations and explicitly include the dependence of the spin exchange interactions on the dynamical lattice positions in order to simulate the dynamics of materials with strong spin-lattice coupling.~\cite{pecharsky2009making, aliev2021giant} For this purpose, spin-lattice dynamics (SLD) simulations add the lattice dynamics and the lattice-dependent spin exchange interactions to the ASD simulations and have been reported to be an effective tool to predict magnetic and thermodynamic properties of magnetic materials more accurately.~\cite{ma2008large,wu2018magnon,hellsvik2019general,perera2017collective} However, SLD simulations have not been applied to analyze the spin and lattice contributions to the magnetocaloric entropy change thus far.

In the current study, we use SLD simulations to quantify the spin and lattice contributions to the magnetocaloric entropy change with a particular focus on understanding the effect of different types of magnetic exchange interactions on MCE. For this purpose, we carry out a thorough comparison between two representative direct and indirect exchange materials, body-centered-cubic (bcc) Fe and hexagonal-closed-pack (hcp) Gd, respectively.~\cite{nolting2009quantum} In magnetic materials with direct exchange interactions, such as bcc Fe, the magnetic exchange interactions are mediated directly by spin-polarized conduction electrons near the Fermi level. In this case, the strength of the direct exchange interaction decreases rapidly with the distance between magnetic ions. In contrast, indirect exchange interactions can couple magnetic moments over relatively large distances.~\cite{parkin1991systematic} Ruderman–Kittel–Kasuya–Yosida (RKKY) interaction is a particular form of indirect magnetic exchange interaction that is dominant in metals with little or no direct overlap between neighboring magnetic electrons.~\cite{ruderman1954indirect}. Instead, the exchange interactions between magnetic ions are mediated by conduction electrons. hcp Gd is an archetypal example of the RKKY interaction.~\cite{lindgaard1975theoretical,hindmarch2003direct,scheie2022spin} The RKKY interaction features an oscillating interaction strength with a periodicity determined by the Fermi wavevector that can lead to longer-range interactions between magnetic ions.~\cite{parkin1990oscillations} With detailed SLD simulations of bcc Fe and hcp Gd, we demonstrate that longer-range RKKY interactions can lead to stronger spin-lattice coupling affecting low-frequency and long-wavelength phonons, which gives rise to a much higher contribution from the lattice to the magnetocaloric entropy change. Our study provides a microscopic mechanism for the enhancement of MCE via spin-lattice coupling and suggests that RKKY interaction is a preferable type of exchange interaction when searching for materials with a strong MCE. We note that, since the electronic contribution to the magnetocaloric entropy change is negligible in hcp Gd,~\cite{martinho2022realistic} the electronic entropy contribution is not discussed in this work.




\section{Computational Methods}
Density functional theory (DFT) calculations were conducted
using the Vienna ab initio simulation package (VASP) based on
the projected augmented wave pseudopotentials.~\cite{kresse1996efficient} The Perdew-Burke-Ernzerhof form of generalized gradient approximation (PBE-GGA)~\cite{blochl1994projector} was used for structural optimization. A plane-wave cut-off energy of 600 eV was utilized for all of our calculations. The energy and force convergence criteria were set to be 1 $\times$ 10$^{-5}$ and 0.01 eV/\AA, respectively. Monkhorst-Pack~\cite{pack1977special} \textbf{k}-point grids of (20 $\times$ 20 $\times$ 20) and (16 $\times$ 16 $\times$ 9) were used to sample the Brillouin zone for the optimization of bcc Fe and hcp Gd, respectively. Magnetic exchange parameters $J_{ij}$ were extracted from a full-potential linear muffin-tin orbital method (FP-LMTO) calculation using the SPR-KKR code,~\cite{ebert2011calculating} in which the spin-configuration-dependent ground-state energy was fitted to a Heisenberg Hamiltonian
\begin{equation}
    H_h = -\sum_{i \neq j} J_{ij} \mathbf{e}_i \cdot \mathbf{e}_j,
\end{equation}
where $\mathbf{e}_i$ and $\mathbf{e}_j$ are unit vectors pointing in the direction of local magnetic moments at atomic site $i$ and $j$.

Spin lattice dynamics (SLD) simulations (See the Supplementary Information for details) were performed using $20 \times 20 \times 20$ supercells using the LAMMPS program.~\cite{plimpton1995fast,tranchida2018massively} Molecular dynamics (MD) timestep, spin, and lattice thermostat damping constants were set to 0.1 fs, 0.1 (Gilbert damping with no units), and 0.1 ps, respectively. The spins were oriented along the $z$-direction at the start of the simulation. To measure the magnetic properties in the canonical ensemble, we initially thermalized the system under NVT dynamics at the target spin and lattice temperatures for 40 ps and then sampled the target properties for 10 ps using a sample interval of 0.001 ps. For pressure-controlled simulations, after the initial 40 ps of temperature equilibration, we froze the spin configuration and ran isobaric-isothermal NPT dynamics to allow the system to thermally expand, while still accounting for the effect of the magnetic pressure generated by the spin Hamiltonian. The pressure damping parameter was set to 10 ps. The pressure equilibration run was terminated once the system pressure dropped below 0.05 GPa. After this, the spin configuration was unfrozen and another equilibration run was carried out under NVT dynamics for 20 ps. Unfreezing the spin configurations causes a small jump in the pressure, typically within the range of +/-2 GPa. To reduce this pressure fluctuation, a series of uniform isotropic box deformations were performed under the NVE ensemble. During this procedure, the box was deformed in 0.02\% increments every 2 ps until the magnitude of the pressure was reduced to negligible values ($<$10 MPa).

\section{Results and Discussions}

\begin{figure}[!htb]
\includegraphics[width=0.75\textwidth]{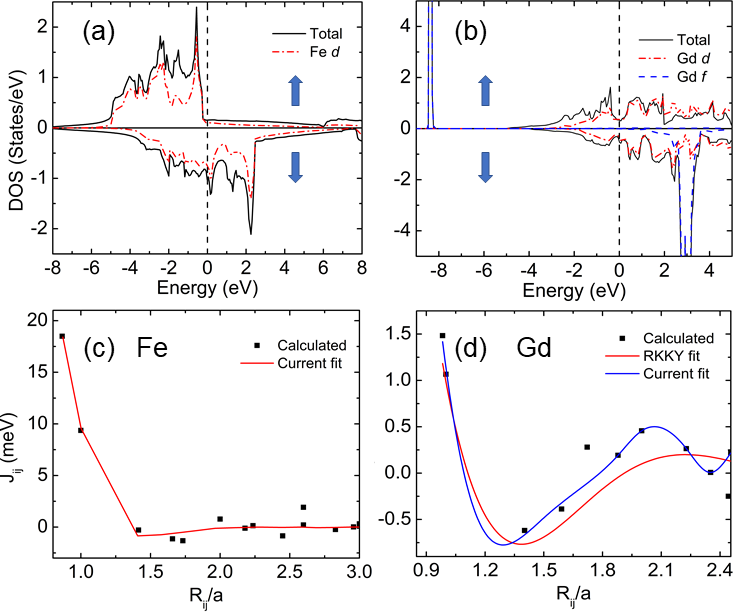}
\caption{\textbf{Comparison of Direct Exchange and RKKY Exchange Interactions in bcc Fe and hcp Gd.} Total and orbital projected density of states for (a) bcc Fe and (b) hcp Gd. The up and down spin channels are indicated with up and down arrows, respectively. The dashed vertical line at 0 eV denotes the Fermi level. The conduction electrons at the Fermi level are spin-polarized in bcc Fe and non-polarized in hcp Gd. Magnetism in hcp Gd is associated with the deep 4$f$ level instead located -8\,eV below the Fermi level. Exchange coupling constants $J_{ij}$ as a function of interatomic distance ($R_{ij}/a$, where $a$ is the lattice constant) between atoms $i$ and $j$ for (c) bcc Fe and (d) hcp Gd. The exchange interaction in bcc Fe decays rapidly and is, thus, short-ranged. In contrast, the exchange interaction in hcp Gd shows an oscillatory behavior and decays more slowly with distance.} 
\label{fig:fig1}
\end{figure}

The spin-dependent electron density of states (DOS) of bcc Fe and hcp Gd is shown in Fig.~\ref{fig:fig1}(a) and (b). In bcc Fe, the conducting $d$ electrons near the Femi level are spin-polarized and are responsible for the direct exchange interactions. In contrast, in hcp Gd, the conducting  electrons near the Fermi level are not spin-polarized while the magnetism comes from the deep $f$ electrons. In this case, the conducting electrons mediate the RKKY interactions. The magnetic exchange parameters ($J_{ij}$) as a function of interatomic distance ($R_{ij}$) for bcc Fe are given in Fig.~\ref{fig:fig1}(c). In bcc Fe, $J_{ij}$ is short-ranged as it decreases rapidly with distance and the values are smaller by at least an order of magnitude after the first two nearest neighbor interactions. Similar dominant contributions within the first two nearest neighbors towards the Heisenberg Hamiltonian for bcc Fe were reported by Wang \textit{et al.}~\cite{wang2010exchange} The calculated exchange parameters have similar values as reported by previous studies.~\cite{wang2010exchange, pajda2001ab, mryasov1996theory, frota2000exchange} The calculated $J_{ij}$ values were fitted to the Bethe-Slater curve to analyze their behavior as a function of interatomic distance. Bethe-Slater curve fitted exchange parameters were previously found to explain the physical properties of bcc Fe accurately.~\cite{kvashnin2016microscopic}


The behavior of $J_{ij}$ as a function of interatomic distance in hcp Gd is quite different as compared to bcc Fe. Despite the fact that Gd is ferromagnetic at a lower temperature, many of the exchange constants are antiferromagnetic. The dependence of the magnetic exchange parameters on the interatomic distance (Fig.~\ref{fig:fig1}(d)) reveals an oscillatory behavior between ferromagnetism and antiferromagnetism as the interatomic distance grows. This is characteristic of the RKKY exchange.~\cite{ruderman1954indirect, kasuya1956theory, yosida1957magnetic} The exchange parameters within the first ten nearest neighbors were considered for further calculations, beyond which the interaction strength becomes negligible. The RKKY-type exchange interaction is mediated by the valence electrons and plays an important role in the magnetic ordering in rare-earth metals or their related compounds, which is sensitive to the atomic separation between these rare-earth atoms. Unlike the transition metals, the large magnetic moment ($\sim$7.5 $\mu_B$ per atom) and strongly correlated behavior of rare-earth hcp Gd originated from half-filled 4$f$ shells. Due to the strong localization of these orbitals, the overlap between neighboring atomic sites is dominated by the 6$s$, 6$p$, and 5$d$ states.~\cite{kurz2002magnetism} The direct exchange contribution from the 4$f$-orbital has been found to be small and antiferromagnetic and hence does not affect the magnetic ordering significantly.~\cite{zhang2017manifestation}The simulated $J_{ij}$ values for hcp Gd agree well with previously reported computational and experimental values.~\cite{turek2006exchange, kvashnin2015exchange,scheie2022spin} The calculated RKKY-type $J_{ij}$ values were fitted with Bathe-Slater curves using different cut-off radii. The details of the fitting process are given in the Supplementary Information.

\begin{figure}[!htb]
\includegraphics[width=0.75\textwidth]{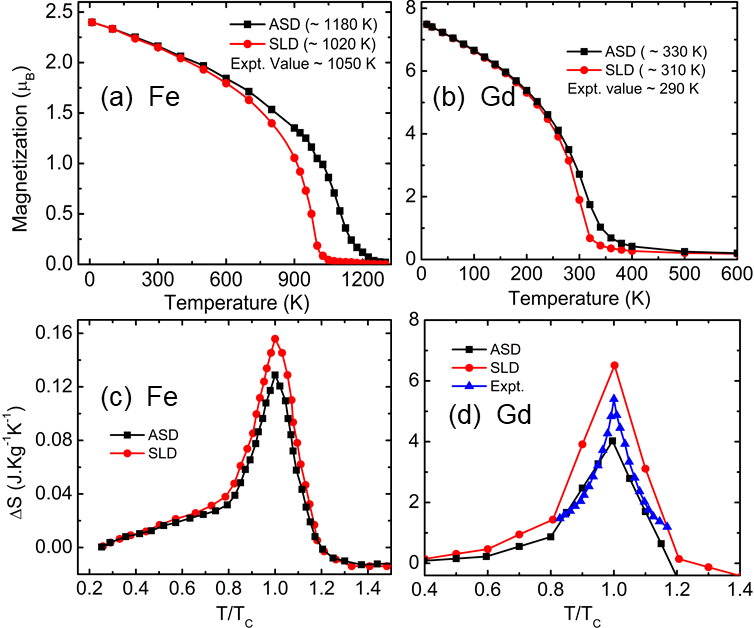}
\caption{\textbf{The ASD and SLD Simulation of the Curie Temperature and Isothermal Entropy Change in bcc Fe and hcp Gd.} The simulated magnetization as a function of temperature is shown for (a) bcc Fe and (b) hcp Gd using both ASD and SLD.  The total isothermal entropy changes ($\Delta S$) are shown for (c) bcc Fe and (d) hcp Gd calculated using both ASD and SLD methods for a field change of 2\,T. The $T_C$ and $\Delta S$ values are provided and compared with the experimentally measured values when available. Including lattice dynamics and spin-lattice coupling leads to a better agreement with experimental values. We did not find available experimental $\Delta S$ data for bcc Fe.} 
\label{fig:fig2}
\end{figure}

The calculated $J_{ij}$ values were used to simulate the Curie temperatures ($T_C$) of bcc Fe and hcp Gd using ASD as well as SLD simulations. The resulting magnetization versus temperature curves are displayed in Fig.~\ref{fig:fig2}(a) and (b). The results were then fitted to a simple power-law decay function of the form $M(T)= (1-\frac{T}{T_C})^\beta$; where $M$ is the magnetization and $\beta$ is the critical exponent. The calculated $T_C$ values with the static-lattice-based ASD approach are 1180\,K and 330\,K for bcc Fe and hcp Gd, respectively, and have a fair agreement with experimental measurements. However, a spin-only model for itinerant magnetism is invalid by construction for quantitative studies as the anomalous temperature dependence of the lattice constants observed in bcc Fe and hcp Gd is completely missing from the atomistic spin dynamics simulations.~\cite{ma2008large} Proper incorporation of finite-temperature lattice dynamics should improve the calculated $T_C$ that is dominated by itinerant $d-$ and $f-$ electron magnetism. The lattice effect in the description of the finite-temperature magnetism in bcc Fe and hcp Gd has been done recently.~\cite{ma2014dynamic} Incorporating the lattice dynamics into effect, our SLD simulations predicted the T$_C$ to be 1020\,K and 310\,K, respectively, for bcc Fe and hcp Gd, with better agreements with the experimentally measured values. The higher difference between ASD and SLD simulated values in bcc Fe indicates a stronger effect of lattice vibrations at the higher Curie temperature.

To understand the difference between the magnetocaloric responses of materials with direct and indirect RKKY exchange coupling parameters, we calculated the isothermal entropy changes with both ASD and SLD simulations for bcc Fe (Fig.~\ref{fig:fig2}(c)) and hcp Gd (Fig.~\ref{fig:fig2}(d)). The external-field-dependent magnetization versus temperature curves were simulated (Figs.~S1 and S3 in the Supplementary Information) and the entropy changes were evaluated using Eqn.~\ref{eqn:entropy_change}. The total entropy change of bcc Fe was calculated to be 0.13 and 0.16\,J\,kg$^{-1}$\, K$^{-1}$ with ASD and SLD simulations, respectively for a magnetic field change of 2\,T. The difference between the entropy change values (0.03\,J\,kg$^{-1}$\,K$^{-1}$) can be attributed to the lattice contribution to the isothermal entropy change at the transition temperature, which amounts to 23\% of the pure spin contribution from the ASD simulation. Next, we analyze the effect of indirect RKKY exchange on the isothermal entropy change in hcp Gd using similar approaches. The evaluated isothermal entropy change using the spin-only Hamiltonian through the ASD simulation is $\sim$4.5\,J\,kg$^{-1}$\,K$^{-1}$ for a magnetic field change of 2\,T. This calculated value based on ASD is smaller compared to the experimentally measured value of $\sim$6\,J\,kg$^{-1}$\,K$^{-1}$.~\cite{gottschall2019making} The sizable discrepancy suggests that the missing lattice dynamics in the ASD simulation can be significant in indirect RKKY exchange materials such as hcp Gd. To verify this hypothesis, we performed SLD simulations of hcp Gd using the interatomic potential developed by Baskes \textit{et al.}~\cite{baskes1994modified} and an isothermal entropy change of $\sim$6.8\,J\,kg$^{-1}$\,K$^{-1}$ was predicted. This result suggests a lattice entropy contribution of 2.3\,J\,kg$^{-1}$\,K$^{-1}$ for a magnetic field change of 2\,T, which is 51\% of the pure spin contribution. Our findings agree well with Martinho Vieira et al.~\cite{martinho2022realistic}, where a Monte Carlo simulation along with DFT was used to study the magnetocaloric response in hcp Gd. The SLD evaluated total entropy change is higher than the measured value by 0.8\,J\,kg$^{-1}$\,K$^{-1}$, which can be potentially attributed to the sample purity and measurement uncertainty in the experiment.

\begin{figure}[!htb]
\includegraphics[width=0.75\textwidth]{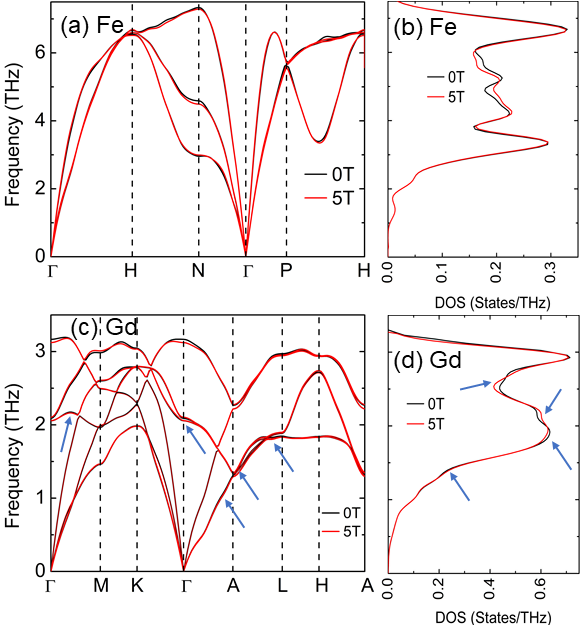}
\caption{\textbf{External-magnetic-field-dependent phonon spectra and density of states (DOS)}. Simulated phonon spectra and DOS in (a, b) bcc Fe and (c, d) hcp Gd at 0\,T and 5\,T with the SLD approach. The changes in low-frequency phonons for hcp Gd are indicated with arrow marks. The magnetic field has a stronger influence on low-frequency and long-wavelength phonons in hcp Gd.} 
\label{fig:fig3}
\end{figure}

The significant lattice entropy change with an applied magnetic field indicates a stronger magnon-phonon coupling in indirect RKKY-exchange-based materials. Microscopically, this result suggests that the phonon structure in hcp Gd near the Curie temperature is sensitively tuned by the external magnetic field. Therefore, it is informative to explicitly examine the phonon dispersion relation of hcp Gd as influenced by an external magnetic field to determine which phonon modes are mostly affected by the applied field. For this purpose, the frequencies of the phonon modes were calculated from solving the dynamic matrix elements obtained from the lattice Green's functions that can be directly calculated from the atomic trajectories in the SLD simulation.~\cite{kong2011phonon}  Figure~\ref{fig:fig3} shows the phonon dispersions of bcc Fe and hcp Gd calculated with magnetic fields of 0\,T and 5\,T, respectively. Data with other applied field values are included in the Supplementary Information (Figs.~S2 and S4). As seen from Fig.~\ref{fig:fig3}(a), in bcc Fe, significant changes in the phonon dispersion can only be noticed at higher phonon frequency ranges ($>$ 3\,THz) and near the Brillouin zone boundaries, whereas the low-frequency and long-wavelength phonons remain unaffected. This feature is more clearly shown in the phonon density of states shown in Fig.~\ref{fig:fig3}(b). This observation can be understood as follows. Since the spin-lattice coupling originates from the dependence of the magnetic exchange parameters on the interatomic distance ($\frac{d J_{ij}}{d R_{ij}}$), the rapidly decreasing $J_{ij}$ as a function of $R_{ij}$ in direct-exchange materials [such as bcc Fe, as shown in Fig.~\ref{fig:fig1}(c)] determines that the spin-lattice coupling mainly affects the short-wavelength lattice vibrations, with a wavelength on the order of the range of the direct exchange. These short-wavelength phonons usually reside near the Brillouin zone boundary and within the higher frequency range, so their occupation is lower at a given temperature, leading to a smaller contribution to the entropy change. In contrast, the phonon dispersion of hcp Gd, as shown in Fig.~\ref{fig:fig3}(c), shows noticeable changes in lower-frequency and longer-wavelength ranges, as labeled by the arrows in Fig.~\ref{fig:fig3}(c), indicating that the spin-lattice coupling in hcp Gd occurs on a larger length scale. This is consistent with the oscillatory behavior of the magnetic exchange parameters as a result of the indirect RKKY exchange interaction, as shown in Fig.~\ref{fig:fig1}(d). Although the overall magnetic exchange strength in hcp Gd is weaker than that in bcc Fe, which leads to a lower Curie temperature in hcp Gd, the much slower decay of the magnetic exchange parameters and their oscillatory behavior as a function of distance leads to significant $\frac{d J_{ij}}{d R_{ij}}$ at longer distances. As a result of this long-range spin-lattice coupling, lattice vibrations associated with phonons with longer wavelengths are more affected by the external field, which can also be seen in the field-dependent phonon density of states shown in Fig.~\ref{fig:fig3}(d). Since these phonons have lower frequencies and, thus, higher occupation numbers at a given temperature, they contribute more to the field-induced isothermal entropy change.  

\begin{figure}[!htb]
\includegraphics[width=0.5\textwidth]{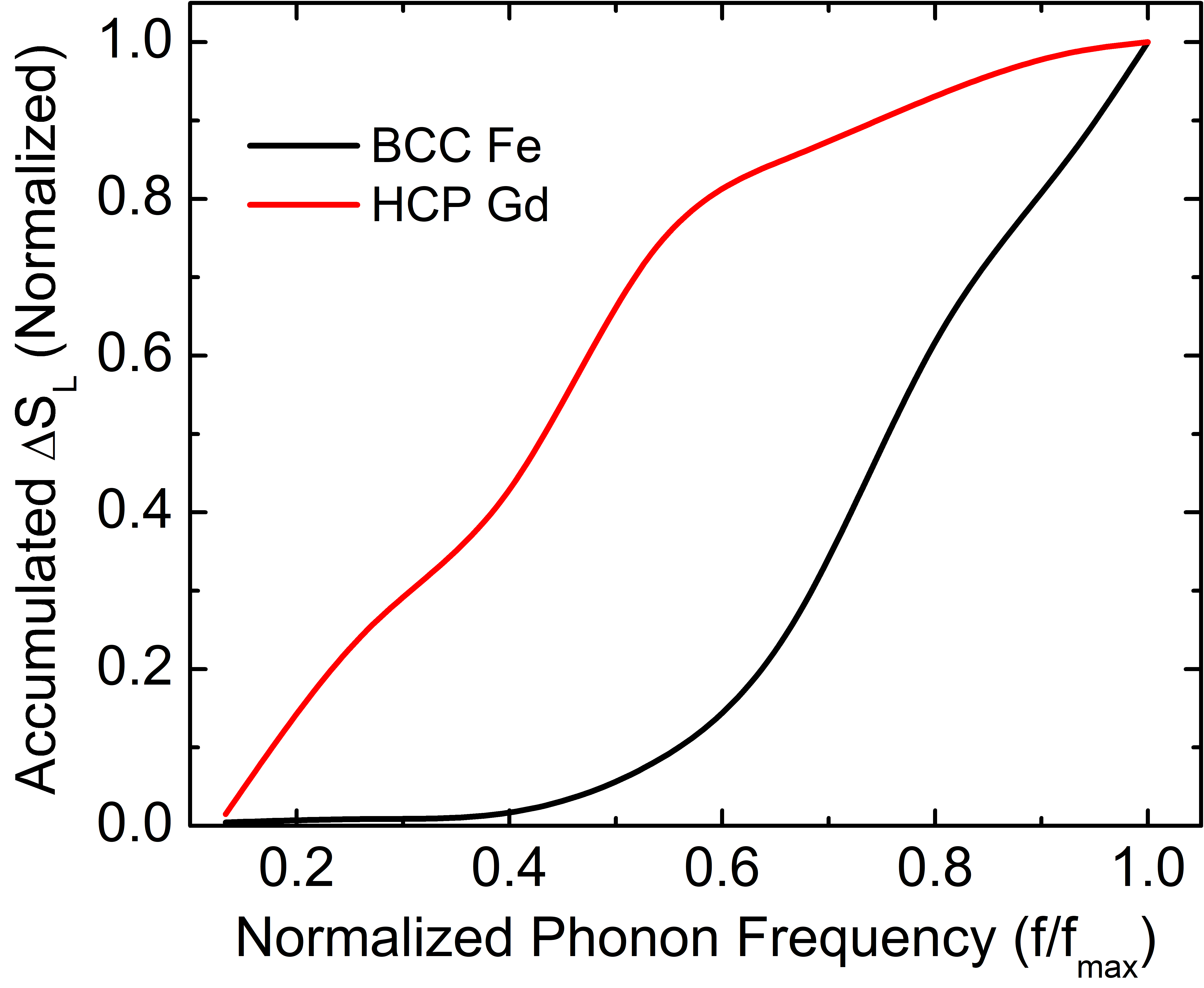}
\caption{\textbf{The accumulated contribution to the total lattice entropy change from phonon modes as a function of phonon frequency in bcc Fe and hcp Gd.} The phonon frequency is normalized to the maximum phonon frequency in either material. The accumulated contribution is also normalized to the total lattice entropy change. The entropy change is evaluated with a magnetic field of 2\,T. It is clearly shown that lower-frequency phonons in hcp Gd has a much more significant contribution than those in bcc Fe.} 
\label{fig:fig4}
\end{figure}

 To compare the lattice entropy contribution from different phonon modes in bcc Fe and hcp Gd more clearly, we further evaluated the lattice entropy change directly based on the field-dependent phonon dispersions. The vibrational entropy of a particular phonon mode with frequency $\omega$ in a harmonic crystal is given by the standard formula for non-interaction bosons:~\cite{wallace2002statistical} 
 \begin{equation}
     S_{ph}(\omega,T)=k_B [(n+1)\ln{(n+1)}-n\ln{n}],
     \label{eqn:phonon_entropy}
 \end{equation}
 where $k_B$ is the Boltzmann constant and $n$ is the occupation number of this phonon mode. Although this result is only rigorously true, it has been shown that,~\cite{hui1975thermodynamics} to the leading order in perturbation theory, Eqn.~\ref{eqn:phonon_entropy} is still valid in anharmonic crystals as long as the renormalized phonon frequencies are used. In our case, the phonon frequencies extracted from the SLD simulation include the full renormalization effect due to both anharmonic phonon-phonon interactions and spin-lattice interactions. 
 Using Eqn.~\ref{eqn:phonon_entropy}, the calculated total lattice entropy change from the field-dependent phonon dispersions for a field change of 2\,T is 0.05\,J\,kg$^{-1}$\,K$^{-1}$ for bcc Fe and 2.5\,J\,kg$^{-1}$\,K$^{-1}$ for hcp Gd. These values are similar to those evaluated by comparing the ASD and SLD simulations, as shown in Fig.~\ref{fig:fig2}(c) and (d). To further contrast the effects of direct exchange and indirect RKKY exchange interactions on lattice entropy changes, the accumulated contribution to the total lattice entropy change from phonons with different frequencies in bcc Fe and hcp Gd under a field change of 2\,T is shown in Fig.~\ref{fig:fig4}. As clearly seen in Fig.~\ref{fig:fig4}, the lower-frequency phonons have negligible contributions toward the lattice entropy change in bcc Fe, whereas in hcp Gd, the lower-frequency phonons have a significant contribution towards the lattice entropy change. This result confirms that the indirect RKKY-type exchange in hcp Gd can impact the short- as well as long-wavelength phonons due to its long interaction range, while the direct exchange in bcc Fe is short-ranged and can only affect the short-wavelength phonons. Our analysis provides a microscopic mechanism of how indirect RKKY exchange can lead to longer-range spin-lattice coupling and a significantly enhanced lattice contribution to the isothermal magnetocaloric entropy change. 
 
\section{Conclusion}
In summary, we applied SLD simulation to directly evaluate the spin and lattice contributions to the isothermal magnetocaloric entropy change in bcc Fe and hcp Gd. Based on a detailed analysis of the field-dependent phonon properties, we conclude that the indirect RKKY-type exchange in hcp Gd leads to a long-range spin-lattice coupling that affects long-wavelength and low-frequency phonons and, thus, causes an enhanced lattice contribution to the total entropy change. Our work provides a microscopic picture of how different types of spin-lattice coupling can give rise to distinct magnetocaloric responses and suggests that indirect RKKY exchange interactions are more desirable for a large MCE response, potentially guiding the future search for more efficient MCE materials.

\begin{acknowledgments}
We thank Dr. Amir Jahromi, Dr. Ali Kashani, and Dr. Leo Ma for their helpful discussions. This work is based on research supported by the National Aeronautics and Space Administration (NASA) under award number 80NSSC21K1812. This work used Stampede2 at Texas Advanced Computing Center (TACC) through allocation MAT200011 from the Advanced Cyberinfrastructure Coordination Ecosystem: Services \& Support (ACCESS) program, which is supported by National Science Foundation grants 2138259, 2138286, 2138307, 2137603, and 2138296. Use was also made of computational facilities purchased with funds from the National Science Foundation (award number CNS-1725797) and administered by the Center for Scientific Computing (CSC) at the University of California, Santa Barbara (UCSB). The CSC is supported by the California NanoSystems Institute and the Materials Research Science and Engineering Center (MRSEC; NSF DMR-1720256) at UCSB. 
\end{acknowledgments}

\bibliography{references.bib}

\begin{thebibliography}{63}%
\makeatletter
\providecommand \@ifxundefined [1]{%
 \@ifx{#1\undefined}
}%
\providecommand \@ifnum [1]{%
 \ifnum #1\expandafter \@firstoftwo
 \else \expandafter \@secondoftwo
 \fi
}%
\providecommand \@ifx [1]{%
 \ifx #1\expandafter \@firstoftwo
 \else \expandafter \@secondoftwo
 \fi
}%
\providecommand \natexlab [1]{#1}%
\providecommand \enquote  [1]{``#1''}%
\providecommand \bibnamefont  [1]{#1}%
\providecommand \bibfnamefont [1]{#1}%
\providecommand \citenamefont [1]{#1}%
\providecommand \href@noop [0]{\@secondoftwo}%
\providecommand \href [0]{\begingroup \@sanitize@url \@href}%
\providecommand \@href[1]{\@@startlink{#1}\@@href}%
\providecommand \@@href[1]{\endgroup#1\@@endlink}%
\providecommand \@sanitize@url [0]{\catcode `\\12\catcode `\$12\catcode
  `\&12\catcode `\#12\catcode `\^12\catcode `\_12\catcode `\%12\relax}%
\providecommand \@@startlink[1]{}%
\providecommand \@@endlink[0]{}%
\providecommand \url  [0]{\begingroup\@sanitize@url \@url }%
\providecommand \@url [1]{\endgroup\@href {#1}{\urlprefix }}%
\providecommand \urlprefix  [0]{URL }%
\providecommand \Eprint [0]{\href }%
\providecommand \doibase [0]{https://doi.org/}%
\providecommand \selectlanguage [0]{\@gobble}%
\providecommand \bibinfo  [0]{\@secondoftwo}%
\providecommand \bibfield  [0]{\@secondoftwo}%
\providecommand \translation [1]{[#1]}%
\providecommand \BibitemOpen [0]{}%
\providecommand \bibitemStop [0]{}%
\providecommand \bibitemNoStop [0]{.\EOS\space}%
\providecommand \EOS [0]{\spacefactor3000\relax}%
\providecommand \BibitemShut  [1]{\csname bibitem#1\endcsname}%
\let\auto@bib@innerbib\@empty
\bibitem [{\citenamefont {Pecharsky}\ and\ \citenamefont
  {Gschneidner~Jr}(1999)}]{pecharsky1999magnetocaloric}%
  \BibitemOpen
  \bibfield  {author} {\bibinfo {author} {\bibfnamefont {V.~K.}\ \bibnamefont
  {Pecharsky}}\ and\ \bibinfo {author} {\bibfnamefont {K.~A.}\ \bibnamefont
  {Gschneidner~Jr}},\ }\bibfield  {title} {\bibinfo {title} {Magnetocaloric
  effect and magnetic refrigeration},\ }\href@noop {} {\bibfield  {journal}
  {\bibinfo  {journal} {Journal of Magnetism and Magnetic Materials}\ }\textbf
  {\bibinfo {volume} {200}},\ \bibinfo {pages} {44} (\bibinfo {year}
  {1999})}\BibitemShut {NoStop}%
\bibitem [{\citenamefont {Balli}\ \emph {et~al.}(2017)\citenamefont {Balli},
  \citenamefont {Jandl}, \citenamefont {Fournier},\ and\ \citenamefont
  {Kedous-Lebouc}}]{balli2017advanced}%
  \BibitemOpen
  \bibfield  {author} {\bibinfo {author} {\bibfnamefont {M.}~\bibnamefont
  {Balli}}, \bibinfo {author} {\bibfnamefont {S.}~\bibnamefont {Jandl}},
  \bibinfo {author} {\bibfnamefont {P.}~\bibnamefont {Fournier}},\ and\
  \bibinfo {author} {\bibfnamefont {A.}~\bibnamefont {Kedous-Lebouc}},\
  }\bibfield  {title} {\bibinfo {title} {Advanced materials for magnetic
  cooling: Fundamentals and practical aspects},\ }\href@noop {} {\bibfield
  {journal} {\bibinfo  {journal} {Appl. Phys. Rev.}\ }\textbf {\bibinfo
  {volume} {4}},\ \bibinfo {pages} {021305} (\bibinfo {year}
  {2017})}\BibitemShut {NoStop}%
\bibitem [{\citenamefont {Shen}\ \emph {et~al.}(2009)\citenamefont {Shen},
  \citenamefont {Sun}, \citenamefont {Hu}, \citenamefont {Zhang},\ and\
  \citenamefont {Cheng}}]{shen2009recent}%
  \BibitemOpen
  \bibfield  {author} {\bibinfo {author} {\bibfnamefont {B.}~\bibnamefont
  {Shen}}, \bibinfo {author} {\bibfnamefont {J.}~\bibnamefont {Sun}}, \bibinfo
  {author} {\bibfnamefont {F.}~\bibnamefont {Hu}}, \bibinfo {author}
  {\bibfnamefont {H.}~\bibnamefont {Zhang}},\ and\ \bibinfo {author}
  {\bibfnamefont {Z.}~\bibnamefont {Cheng}},\ }\bibfield  {title} {\bibinfo
  {title} {Recent progress in exploring magnetocaloric materials},\ }\href@noop
  {} {\bibfield  {journal} {\bibinfo  {journal} {Adv. Mater.}\ }\textbf
  {\bibinfo {volume} {21}},\ \bibinfo {pages} {4545} (\bibinfo {year}
  {2009})}\BibitemShut {NoStop}%
\bibitem [{\citenamefont {Franco}\ \emph {et~al.}(2018)\citenamefont {Franco},
  \citenamefont {Bl{\'a}zquez}, \citenamefont {Ipus}, \citenamefont {Law},
  \citenamefont {Moreno-Ram{\'\i}rez},\ and\ \citenamefont
  {Conde}}]{franco2018magnetocaloric}%
  \BibitemOpen
  \bibfield  {author} {\bibinfo {author} {\bibfnamefont {V.}~\bibnamefont
  {Franco}}, \bibinfo {author} {\bibfnamefont {J.}~\bibnamefont
  {Bl{\'a}zquez}}, \bibinfo {author} {\bibfnamefont {J.}~\bibnamefont {Ipus}},
  \bibinfo {author} {\bibfnamefont {J.}~\bibnamefont {Law}}, \bibinfo {author}
  {\bibfnamefont {L.}~\bibnamefont {Moreno-Ram{\'\i}rez}},\ and\ \bibinfo
  {author} {\bibfnamefont {A.}~\bibnamefont {Conde}},\ }\bibfield  {title}
  {\bibinfo {title} {Magnetocaloric effect: From materials research to
  refrigeration devices},\ }\href@noop {} {\bibfield  {journal} {\bibinfo
  {journal} {Prog. Mater. Sci.}\ }\textbf {\bibinfo {volume} {93}},\ \bibinfo
  {pages} {112} (\bibinfo {year} {2018})}\BibitemShut {NoStop}%
\bibitem [{\citenamefont {Balz}\ \emph {et~al.}(2019)\citenamefont {Balz},
  \citenamefont {Lampen-Kelley}, \citenamefont {Banerjee}, \citenamefont {Yan},
  \citenamefont {Lu}, \citenamefont {Hu}, \citenamefont {Yadav}, \citenamefont
  {Takano}, \citenamefont {Liu}, \citenamefont {Tennant} \emph
  {et~al.}}]{balz2019finite}%
  \BibitemOpen
  \bibfield  {author} {\bibinfo {author} {\bibfnamefont {C.}~\bibnamefont
  {Balz}}, \bibinfo {author} {\bibfnamefont {P.}~\bibnamefont {Lampen-Kelley}},
  \bibinfo {author} {\bibfnamefont {A.}~\bibnamefont {Banerjee}}, \bibinfo
  {author} {\bibfnamefont {J.}~\bibnamefont {Yan}}, \bibinfo {author}
  {\bibfnamefont {Z.}~\bibnamefont {Lu}}, \bibinfo {author} {\bibfnamefont
  {X.}~\bibnamefont {Hu}}, \bibinfo {author} {\bibfnamefont {S.~M.}\
  \bibnamefont {Yadav}}, \bibinfo {author} {\bibfnamefont {Y.}~\bibnamefont
  {Takano}}, \bibinfo {author} {\bibfnamefont {Y.}~\bibnamefont {Liu}},
  \bibinfo {author} {\bibfnamefont {D.~A.}\ \bibnamefont {Tennant}}, \emph
  {et~al.},\ }\bibfield  {title} {\bibinfo {title} {Finite field regime for a
  quantum spin liquid in $\alpha$- rucl 3},\ }\href@noop {} {\bibfield
  {journal} {\bibinfo  {journal} {Physical Review B}\ }\textbf {\bibinfo
  {volume} {100}},\ \bibinfo {pages} {060405} (\bibinfo {year}
  {2019})}\BibitemShut {NoStop}%
\bibitem [{\citenamefont {Czajka}\ \emph {et~al.}(2021)\citenamefont {Czajka},
  \citenamefont {Gao}, \citenamefont {Hirschberger}, \citenamefont
  {Lampen-Kelley}, \citenamefont {Banerjee}, \citenamefont {Yan}, \citenamefont
  {Mandrus}, \citenamefont {Nagler},\ and\ \citenamefont
  {Ong}}]{czajka2021oscillations}%
  \BibitemOpen
  \bibfield  {author} {\bibinfo {author} {\bibfnamefont {P.}~\bibnamefont
  {Czajka}}, \bibinfo {author} {\bibfnamefont {T.}~\bibnamefont {Gao}},
  \bibinfo {author} {\bibfnamefont {M.}~\bibnamefont {Hirschberger}}, \bibinfo
  {author} {\bibfnamefont {P.}~\bibnamefont {Lampen-Kelley}}, \bibinfo {author}
  {\bibfnamefont {A.}~\bibnamefont {Banerjee}}, \bibinfo {author}
  {\bibfnamefont {J.}~\bibnamefont {Yan}}, \bibinfo {author} {\bibfnamefont
  {D.~G.}\ \bibnamefont {Mandrus}}, \bibinfo {author} {\bibfnamefont {S.~E.}\
  \bibnamefont {Nagler}},\ and\ \bibinfo {author} {\bibfnamefont
  {N.}~\bibnamefont {Ong}},\ }\bibfield  {title} {\bibinfo {title}
  {Oscillations of the thermal conductivity in the spin-liquid state of
  $\alpha$-{RuCl$_3$}},\ }\href@noop {} {\bibfield  {journal} {\bibinfo
  {journal} {Nature Physics}\ }\textbf {\bibinfo {volume} {17}},\ \bibinfo
  {pages} {915} (\bibinfo {year} {2021})}\BibitemShut {NoStop}%
\bibitem [{\citenamefont {Shirron}\ \emph {et~al.}(2012)\citenamefont
  {Shirron}, \citenamefont {Kimball}, \citenamefont {Fixsen}, \citenamefont
  {Kogut}, \citenamefont {Li},\ and\ \citenamefont
  {DiPirro}}]{shirron2012design}%
  \BibitemOpen
  \bibfield  {author} {\bibinfo {author} {\bibfnamefont {P.~J.}\ \bibnamefont
  {Shirron}}, \bibinfo {author} {\bibfnamefont {M.~O.}\ \bibnamefont
  {Kimball}}, \bibinfo {author} {\bibfnamefont {D.~J.}\ \bibnamefont {Fixsen}},
  \bibinfo {author} {\bibfnamefont {A.~J.}\ \bibnamefont {Kogut}}, \bibinfo
  {author} {\bibfnamefont {X.}~\bibnamefont {Li}},\ and\ \bibinfo {author}
  {\bibfnamefont {M.~J.}\ \bibnamefont {DiPirro}},\ }\bibfield  {title}
  {\bibinfo {title} {Design of the {PIXIE} adiabatic demagnetization
  refrigerators},\ }\href@noop {} {\bibfield  {journal} {\bibinfo  {journal}
  {Cryogenics}\ }\textbf {\bibinfo {volume} {52}},\ \bibinfo {pages} {140}
  (\bibinfo {year} {2012})}\BibitemShut {NoStop}%
\bibitem [{\citenamefont {Britt}\ and\ \citenamefont
  {Richards}(1981)}]{britt1981adiabatic}%
  \BibitemOpen
  \bibfield  {author} {\bibinfo {author} {\bibfnamefont {R.~D.}\ \bibnamefont
  {Britt}}\ and\ \bibinfo {author} {\bibfnamefont {P.}~\bibnamefont
  {Richards}},\ }\bibfield  {title} {\bibinfo {title} {An adiabatic
  demagnetization refrigerator for infrared bolometers},\ }\href@noop {}
  {\bibfield  {journal} {\bibinfo  {journal} {International Journal of Infrared
  and Millimeter Waves}\ }\textbf {\bibinfo {volume} {2}},\ \bibinfo {pages}
  {1083} (\bibinfo {year} {1981})}\BibitemShut {NoStop}%
\bibitem [{\citenamefont {Sato}\ \emph {et~al.}(2016)\citenamefont {Sato},
  \citenamefont {Okuyama},\ and\ \citenamefont {Kimura}}]{sato2016tiny}%
  \BibitemOpen
  \bibfield  {author} {\bibinfo {author} {\bibfnamefont {T.~J.}\ \bibnamefont
  {Sato}}, \bibinfo {author} {\bibfnamefont {D.}~\bibnamefont {Okuyama}},\ and\
  \bibinfo {author} {\bibfnamefont {H.}~\bibnamefont {Kimura}},\ }\bibfield
  {title} {\bibinfo {title} {Tiny adiabatic-demagnetization refrigerator for a
  commercial superconducting quantum interference device magnetometer},\
  }\href@noop {} {\bibfield  {journal} {\bibinfo  {journal} {Review of
  Scientific Instruments}\ }\textbf {\bibinfo {volume} {87}},\ \bibinfo {pages}
  {123905} (\bibinfo {year} {2016})}\BibitemShut {NoStop}%
\bibitem [{\citenamefont {Aprea}\ \emph {et~al.}(2015)\citenamefont {Aprea},
  \citenamefont {Greco}, \citenamefont {Maiorino},\ and\ \citenamefont
  {Masselli}}]{aprea2015magnetic}%
  \BibitemOpen
  \bibfield  {author} {\bibinfo {author} {\bibfnamefont {C.}~\bibnamefont
  {Aprea}}, \bibinfo {author} {\bibfnamefont {A.}~\bibnamefont {Greco}},
  \bibinfo {author} {\bibfnamefont {A.}~\bibnamefont {Maiorino}},\ and\
  \bibinfo {author} {\bibfnamefont {C.}~\bibnamefont {Masselli}},\ }\bibfield
  {title} {\bibinfo {title} {Magnetic refrigeration: an eco-friendly technology
  for the refrigeration at room temperature},\ }in\ \href@noop {} {\emph
  {\bibinfo {booktitle} {Journal of Physics: Conference Series}}},\ Vol.\
  \bibinfo {volume} {655}\ (\bibinfo {organization} {IOP Publishing},\ \bibinfo
  {year} {2015})\ p.\ \bibinfo {pages} {012026}\BibitemShut {NoStop}%
\bibitem [{\citenamefont {Pecharsky}\ and\ \citenamefont
  {Gschneidner~Jr}(1997)}]{pecharsky1997giant}%
  \BibitemOpen
  \bibfield  {author} {\bibinfo {author} {\bibfnamefont {V.~K.}\ \bibnamefont
  {Pecharsky}}\ and\ \bibinfo {author} {\bibfnamefont {K.~A.}\ \bibnamefont
  {Gschneidner~Jr}},\ }\bibfield  {title} {\bibinfo {title} {Giant
  magnetocaloric effect in {G}d$_5$({S}i$_2${G}e$_2$)},\ }\href@noop {}
  {\bibfield  {journal} {\bibinfo  {journal} {Phys. Rev. Lett.}\ }\textbf
  {\bibinfo {volume} {78}},\ \bibinfo {pages} {4494} (\bibinfo {year}
  {1997})}\BibitemShut {NoStop}%
\bibitem [{\citenamefont {Choe}\ \emph {et~al.}(2003)\citenamefont {Choe},
  \citenamefont {Miller}, \citenamefont {Meyers}, \citenamefont {Chumbley},\
  and\ \citenamefont {Pecharsky}}]{choe2003nanoscale}%
  \BibitemOpen
  \bibfield  {author} {\bibinfo {author} {\bibfnamefont {W.}~\bibnamefont
  {Choe}}, \bibinfo {author} {\bibfnamefont {G.~J.}\ \bibnamefont {Miller}},
  \bibinfo {author} {\bibfnamefont {J.}~\bibnamefont {Meyers}}, \bibinfo
  {author} {\bibfnamefont {S.}~\bibnamefont {Chumbley}},\ and\ \bibinfo
  {author} {\bibfnamefont {A.}~\bibnamefont {Pecharsky}},\ }\bibfield  {title}
  {\bibinfo {title} {{“Nanoscale zippers”} in the crystalline solid.
  {S}tructural variations in the giant magnetocaloric material
  {G}d$_5${S}i$_{1.5}${G}e$_{2.5}$},\ }\href@noop {} {\bibfield  {journal}
  {\bibinfo  {journal} {Chem. Mater.}\ }\textbf {\bibinfo {volume} {15}},\
  \bibinfo {pages} {1413} (\bibinfo {year} {2003})}\BibitemShut {NoStop}%
\bibitem [{\citenamefont {Magen}\ \emph {et~al.}(2003)\citenamefont {Magen},
  \citenamefont {Arnold}, \citenamefont {Morellon}, \citenamefont {Skorokhod},
  \citenamefont {Algarabel}, \citenamefont {Ibarra},\ and\ \citenamefont
  {Kamarad}}]{magen2003pressure}%
  \BibitemOpen
  \bibfield  {author} {\bibinfo {author} {\bibfnamefont {C.}~\bibnamefont
  {Magen}}, \bibinfo {author} {\bibfnamefont {Z.}~\bibnamefont {Arnold}},
  \bibinfo {author} {\bibfnamefont {L.}~\bibnamefont {Morellon}}, \bibinfo
  {author} {\bibfnamefont {Y.}~\bibnamefont {Skorokhod}}, \bibinfo {author}
  {\bibfnamefont {P.}~\bibnamefont {Algarabel}}, \bibinfo {author}
  {\bibfnamefont {M.}~\bibnamefont {Ibarra}},\ and\ \bibinfo {author}
  {\bibfnamefont {J.}~\bibnamefont {Kamarad}},\ }\bibfield  {title} {\bibinfo
  {title} {Pressure-induced three-dimensional ferromagnetic correlations in the
  giant magnetocaloric compound {G}d$_5${G}e$_4$},\ }\href@noop {} {\bibfield
  {journal} {\bibinfo  {journal} {Phys. Rev. Lett.}\ }\textbf {\bibinfo
  {volume} {91}},\ \bibinfo {pages} {207202} (\bibinfo {year}
  {2003})}\BibitemShut {NoStop}%
\bibitem [{\citenamefont {Wada}\ and\ \citenamefont
  {Tanabe}(2001)}]{wada2001giant}%
  \BibitemOpen
  \bibfield  {author} {\bibinfo {author} {\bibfnamefont {H.}~\bibnamefont
  {Wada}}\ and\ \bibinfo {author} {\bibfnamefont {Y.}~\bibnamefont {Tanabe}},\
  }\bibfield  {title} {\bibinfo {title} {Giant magnetocaloric effect of
  {M}n{A}s$_{1-x}${S}b$_x$},\ }\href@noop {} {\bibfield  {journal} {\bibinfo
  {journal} {Appl. Phys. Lett.}\ }\textbf {\bibinfo {volume} {79}},\ \bibinfo
  {pages} {3302} (\bibinfo {year} {2001})}\BibitemShut {NoStop}%
\bibitem [{\citenamefont {Shao}\ \emph {et~al.}(2022)\citenamefont {Shao},
  \citenamefont {Miao}, \citenamefont {Zhang}, \citenamefont {Xu},\ and\
  \citenamefont {Liu}}]{shao2022depth}%
  \BibitemOpen
  \bibfield  {author} {\bibinfo {author} {\bibfnamefont {Y.}~\bibnamefont
  {Shao}}, \bibinfo {author} {\bibfnamefont {X.}~\bibnamefont {Miao}}, \bibinfo
  {author} {\bibfnamefont {Y.}~\bibnamefont {Zhang}}, \bibinfo {author}
  {\bibfnamefont {F.}~\bibnamefont {Xu}},\ and\ \bibinfo {author}
  {\bibfnamefont {J.}~\bibnamefont {Liu}},\ }\bibfield  {title} {\bibinfo
  {title} {An in-depth review of {L}a-{F}e-{S}i magnetocaloric composites:
  structure design and performance enhancement},\ }\href@noop {} {\bibfield
  {journal} {\bibinfo  {journal} {J. Magn. Magn. Mater.}\ }\textbf {\bibinfo
  {volume} {564}},\ \bibinfo {pages} {170057} (\bibinfo {year}
  {2022})}\BibitemShut {NoStop}%
\bibitem [{\citenamefont {Bao}\ \emph {et~al.}(2012)\citenamefont {Bao},
  \citenamefont {Hu}, \citenamefont {Chen}, \citenamefont {Wang}, \citenamefont
  {Sun},\ and\ \citenamefont {Shen}}]{bao2012magnetocaloric}%
  \BibitemOpen
  \bibfield  {author} {\bibinfo {author} {\bibfnamefont {L.}~\bibnamefont
  {Bao}}, \bibinfo {author} {\bibfnamefont {F.}~\bibnamefont {Hu}}, \bibinfo
  {author} {\bibfnamefont {L.}~\bibnamefont {Chen}}, \bibinfo {author}
  {\bibfnamefont {J.}~\bibnamefont {Wang}}, \bibinfo {author} {\bibfnamefont
  {J.}~\bibnamefont {Sun}},\ and\ \bibinfo {author} {\bibfnamefont
  {B.}~\bibnamefont {Shen}},\ }\bibfield  {title} {\bibinfo {title}
  {Magnetocaloric properties of {L}a({F}e, {S}i)$_{13}$-based material and its
  hydride prepared by industrial mischmetal},\ }\href@noop {} {\bibfield
  {journal} {\bibinfo  {journal} {Appl. Phys. Lett.}\ }\textbf {\bibinfo
  {volume} {101}},\ \bibinfo {pages} {162406} (\bibinfo {year}
  {2012})}\BibitemShut {NoStop}%
\bibitem [{\citenamefont {Tegus}\ \emph {et~al.}(2002)\citenamefont {Tegus},
  \citenamefont {Br{\"u}ck}, \citenamefont {Buschow},\ and\ \citenamefont
  {De~Boer}}]{tegus2002transition}%
  \BibitemOpen
  \bibfield  {author} {\bibinfo {author} {\bibfnamefont {O.}~\bibnamefont
  {Tegus}}, \bibinfo {author} {\bibfnamefont {E.}~\bibnamefont {Br{\"u}ck}},
  \bibinfo {author} {\bibfnamefont {K.}~\bibnamefont {Buschow}},\ and\ \bibinfo
  {author} {\bibfnamefont {F.}~\bibnamefont {De~Boer}},\ }\bibfield  {title}
  {\bibinfo {title} {Transition-metal-based magnetic refrigerants for
  room-temperature applications},\ }\href@noop {} {\bibfield  {journal}
  {\bibinfo  {journal} {Nature}\ }\textbf {\bibinfo {volume} {415}},\ \bibinfo
  {pages} {150} (\bibinfo {year} {2002})}\BibitemShut {NoStop}%
\bibitem [{\citenamefont {Krenke}\ \emph {et~al.}(2005)\citenamefont {Krenke},
  \citenamefont {Duman}, \citenamefont {Acet}, \citenamefont {Wassermann},
  \citenamefont {Moya}, \citenamefont {Ma{\~n}osa},\ and\ \citenamefont
  {Planes}}]{krenke2005inverse}%
  \BibitemOpen
  \bibfield  {author} {\bibinfo {author} {\bibfnamefont {T.}~\bibnamefont
  {Krenke}}, \bibinfo {author} {\bibfnamefont {E.}~\bibnamefont {Duman}},
  \bibinfo {author} {\bibfnamefont {M.}~\bibnamefont {Acet}}, \bibinfo {author}
  {\bibfnamefont {E.~F.}\ \bibnamefont {Wassermann}}, \bibinfo {author}
  {\bibfnamefont {X.}~\bibnamefont {Moya}}, \bibinfo {author} {\bibfnamefont
  {L.}~\bibnamefont {Ma{\~n}osa}},\ and\ \bibinfo {author} {\bibfnamefont
  {A.}~\bibnamefont {Planes}},\ }\bibfield  {title} {\bibinfo {title} {Inverse
  magnetocaloric effect in ferromagnetic {N}i--{M}n--{S}n alloys},\ }\href@noop
  {} {\bibfield  {journal} {\bibinfo  {journal} {Nat. Mater.}\ }\textbf
  {\bibinfo {volume} {4}},\ \bibinfo {pages} {450} (\bibinfo {year}
  {2005})}\BibitemShut {NoStop}%
\bibitem [{\citenamefont {Krenke}\ \emph {et~al.}(2007)\citenamefont {Krenke},
  \citenamefont {Duman}, \citenamefont {Acet}, \citenamefont {Wassermann},
  \citenamefont {Moya}, \citenamefont {Ma{\~n}osa}, \citenamefont {Planes},
  \citenamefont {Suard},\ and\ \citenamefont {Ouladdiaf}}]{krenke2007magnetic}%
  \BibitemOpen
  \bibfield  {author} {\bibinfo {author} {\bibfnamefont {T.}~\bibnamefont
  {Krenke}}, \bibinfo {author} {\bibfnamefont {E.}~\bibnamefont {Duman}},
  \bibinfo {author} {\bibfnamefont {M.}~\bibnamefont {Acet}}, \bibinfo {author}
  {\bibfnamefont {E.~F.}\ \bibnamefont {Wassermann}}, \bibinfo {author}
  {\bibfnamefont {X.}~\bibnamefont {Moya}}, \bibinfo {author} {\bibfnamefont
  {L.}~\bibnamefont {Ma{\~n}osa}}, \bibinfo {author} {\bibfnamefont
  {A.}~\bibnamefont {Planes}}, \bibinfo {author} {\bibfnamefont
  {E.}~\bibnamefont {Suard}},\ and\ \bibinfo {author} {\bibfnamefont
  {B.}~\bibnamefont {Ouladdiaf}},\ }\bibfield  {title} {\bibinfo {title}
  {Magnetic superelasticity and inverse magnetocaloric effect in
  {N}i-{M}n-{I}n},\ }\href@noop {} {\bibfield  {journal} {\bibinfo  {journal}
  {Phys. Rev. B}\ }\textbf {\bibinfo {volume} {75}},\ \bibinfo {pages} {104414}
  (\bibinfo {year} {2007})}\BibitemShut {NoStop}%
\bibitem [{\citenamefont {Pecharsky}\ \emph {et~al.}(2003)\citenamefont
  {Pecharsky}, \citenamefont {Holm}, \citenamefont {Gschneidner~Jr},\ and\
  \citenamefont {Rink}}]{pecharsky2003massive}%
  \BibitemOpen
  \bibfield  {author} {\bibinfo {author} {\bibfnamefont {V.}~\bibnamefont
  {Pecharsky}}, \bibinfo {author} {\bibfnamefont {A.}~\bibnamefont {Holm}},
  \bibinfo {author} {\bibfnamefont {K.}~\bibnamefont {Gschneidner~Jr}},\ and\
  \bibinfo {author} {\bibfnamefont {R.}~\bibnamefont {Rink}},\ }\bibfield
  {title} {\bibinfo {title} {Massive magnetic-field-induced structural
  transformation in {G}d$_5${G}e$_4$ and the nature of the giant magnetocaloric
  effect},\ }\href@noop {} {\bibfield  {journal} {\bibinfo  {journal} {Phys.
  Rev. Lett.}\ }\textbf {\bibinfo {volume} {91}},\ \bibinfo {pages} {197204}
  (\bibinfo {year} {2003})}\BibitemShut {NoStop}%
\bibitem [{\citenamefont {Morellon}\ \emph {et~al.}(2004)\citenamefont
  {Morellon}, \citenamefont {Arnold}, \citenamefont {Magen}, \citenamefont
  {Ritter}, \citenamefont {Prokhnenko}, \citenamefont {Skorokhod},
  \citenamefont {Algarabel}, \citenamefont {Ibarra},\ and\ \citenamefont
  {Kamarad}}]{morellon2004pressure}%
  \BibitemOpen
  \bibfield  {author} {\bibinfo {author} {\bibfnamefont {L.}~\bibnamefont
  {Morellon}}, \bibinfo {author} {\bibfnamefont {Z.}~\bibnamefont {Arnold}},
  \bibinfo {author} {\bibfnamefont {C.}~\bibnamefont {Magen}}, \bibinfo
  {author} {\bibfnamefont {C.}~\bibnamefont {Ritter}}, \bibinfo {author}
  {\bibfnamefont {O.}~\bibnamefont {Prokhnenko}}, \bibinfo {author}
  {\bibfnamefont {Y.}~\bibnamefont {Skorokhod}}, \bibinfo {author}
  {\bibfnamefont {P.}~\bibnamefont {Algarabel}}, \bibinfo {author}
  {\bibfnamefont {M.}~\bibnamefont {Ibarra}},\ and\ \bibinfo {author}
  {\bibfnamefont {J.}~\bibnamefont {Kamarad}},\ }\bibfield  {title} {\bibinfo
  {title} {Pressure enhancement of the giant magnetocaloric effect in
  {T}b$_5${S}i$_2${G}e$_2$},\ }\href@noop {} {\bibfield  {journal} {\bibinfo
  {journal} {Phys. Rev. Lett.}\ }\textbf {\bibinfo {volume} {93}},\ \bibinfo
  {pages} {137201} (\bibinfo {year} {2004})}\BibitemShut {NoStop}%
\bibitem [{\citenamefont {Wada}\ \emph {et~al.}(2009)\citenamefont {Wada},
  \citenamefont {Matsuo},\ and\ \citenamefont {Mitsuda}}]{wada2009pressure}%
  \BibitemOpen
  \bibfield  {author} {\bibinfo {author} {\bibfnamefont {H.}~\bibnamefont
  {Wada}}, \bibinfo {author} {\bibfnamefont {S.}~\bibnamefont {Matsuo}},\ and\
  \bibinfo {author} {\bibfnamefont {A.}~\bibnamefont {Mitsuda}},\ }\bibfield
  {title} {\bibinfo {title} {Pressure dependence of magnetic entropy change and
  magnetic transition in {M}n{A}s$_{1-x}${S}b$_x$},\ }\href@noop {} {\bibfield
  {journal} {\bibinfo  {journal} {Phys. Rev. B}\ }\textbf {\bibinfo {volume}
  {79}},\ \bibinfo {pages} {092407} (\bibinfo {year} {2009})}\BibitemShut
  {NoStop}%
\bibitem [{\citenamefont {Kuz’Min}(2007)}]{kuz2007factors}%
  \BibitemOpen
  \bibfield  {author} {\bibinfo {author} {\bibfnamefont {M.}~\bibnamefont
  {Kuz’Min}},\ }\bibfield  {title} {\bibinfo {title} {Factors limiting the
  operation frequency of magnetic refrigerators},\ }\href@noop {} {\bibfield
  {journal} {\bibinfo  {journal} {Appl. Phys, Lett.}\ }\textbf {\bibinfo
  {volume} {90}},\ \bibinfo {pages} {251916} (\bibinfo {year}
  {2007})}\BibitemShut {NoStop}%
\bibitem [{\citenamefont {Bocarsly}\ \emph {et~al.}(2017)\citenamefont
  {Bocarsly}, \citenamefont {Levin}, \citenamefont {Garcia}, \citenamefont
  {Schwennicke}, \citenamefont {Wilson},\ and\ \citenamefont
  {Seshadri}}]{bocarsly2017simple}%
  \BibitemOpen
  \bibfield  {author} {\bibinfo {author} {\bibfnamefont {J.~D.}\ \bibnamefont
  {Bocarsly}}, \bibinfo {author} {\bibfnamefont {E.~E.}\ \bibnamefont {Levin}},
  \bibinfo {author} {\bibfnamefont {C.~A.}\ \bibnamefont {Garcia}}, \bibinfo
  {author} {\bibfnamefont {K.}~\bibnamefont {Schwennicke}}, \bibinfo {author}
  {\bibfnamefont {S.~D.}\ \bibnamefont {Wilson}},\ and\ \bibinfo {author}
  {\bibfnamefont {R.}~\bibnamefont {Seshadri}},\ }\bibfield  {title} {\bibinfo
  {title} {A simple computational proxy for screening magnetocaloric
  compounds},\ }\href@noop {} {\bibfield  {journal} {\bibinfo  {journal} {Chem.
  Mater.}\ }\textbf {\bibinfo {volume} {29}},\ \bibinfo {pages} {1613}
  (\bibinfo {year} {2017})}\BibitemShut {NoStop}%
\bibitem [{\citenamefont {Antropov}\ \emph {et~al.}(1995)\citenamefont
  {Antropov}, \citenamefont {Katsnelson}, \citenamefont {Van~Schilfgaarde},\
  and\ \citenamefont {Harmon}}]{antropov1995ab}%
  \BibitemOpen
  \bibfield  {author} {\bibinfo {author} {\bibfnamefont {V.~P.}\ \bibnamefont
  {Antropov}}, \bibinfo {author} {\bibfnamefont {M.}~\bibnamefont
  {Katsnelson}}, \bibinfo {author} {\bibfnamefont {M.}~\bibnamefont
  {Van~Schilfgaarde}},\ and\ \bibinfo {author} {\bibfnamefont {B.}~\bibnamefont
  {Harmon}},\ }\bibfield  {title} {\bibinfo {title} {Ab initio spin dynamics in
  magnets},\ }\href@noop {} {\bibfield  {journal} {\bibinfo  {journal} {Phys.
  Rev. Lett}\ }\textbf {\bibinfo {volume} {75}},\ \bibinfo {pages} {729}
  (\bibinfo {year} {1995})}\BibitemShut {NoStop}%
\bibitem [{\citenamefont {Skubic}\ \emph {et~al.}(2008)\citenamefont {Skubic},
  \citenamefont {Hellsvik}, \citenamefont {Nordstr{\"o}m},\ and\ \citenamefont
  {Eriksson}}]{skubic2008method}%
  \BibitemOpen
  \bibfield  {author} {\bibinfo {author} {\bibfnamefont {B.}~\bibnamefont
  {Skubic}}, \bibinfo {author} {\bibfnamefont {J.}~\bibnamefont {Hellsvik}},
  \bibinfo {author} {\bibfnamefont {L.}~\bibnamefont {Nordstr{\"o}m}},\ and\
  \bibinfo {author} {\bibfnamefont {O.}~\bibnamefont {Eriksson}},\ }\bibfield
  {title} {\bibinfo {title} {A method for atomistic spin dynamics simulations:
  {i}mplementation and examples},\ }\href@noop {} {\bibfield  {journal}
  {\bibinfo  {journal} {J. Phys. Cond. Matt.}\ }\textbf {\bibinfo {volume}
  {20}},\ \bibinfo {pages} {315203} (\bibinfo {year} {2008})}\BibitemShut
  {NoStop}%
\bibitem [{\citenamefont {Ma}\ and\ \citenamefont
  {Dudarev}(2014)}]{ma2014dynamic}%
  \BibitemOpen
  \bibfield  {author} {\bibinfo {author} {\bibfnamefont {P.-W.}\ \bibnamefont
  {Ma}}\ and\ \bibinfo {author} {\bibfnamefont {S.}~\bibnamefont {Dudarev}},\
  }\bibfield  {title} {\bibinfo {title} {Dynamic magnetocaloric effect in bcc
  iron and hcp gadolinium},\ }\href@noop {} {\bibfield  {journal} {\bibinfo
  {journal} {Physical Review B}\ }\textbf {\bibinfo {volume} {90}},\ \bibinfo
  {pages} {024425} (\bibinfo {year} {2014})}\BibitemShut {NoStop}%
\bibitem [{\citenamefont {Pecharsky}\ \emph {et~al.}(2009)\citenamefont
  {Pecharsky}, \citenamefont {Gschneidner~Jr}, \citenamefont {Mudryk},\ and\
  \citenamefont {Paudyal}}]{pecharsky2009making}%
  \BibitemOpen
  \bibfield  {author} {\bibinfo {author} {\bibfnamefont {V.}~\bibnamefont
  {Pecharsky}}, \bibinfo {author} {\bibfnamefont {K.}~\bibnamefont
  {Gschneidner~Jr}}, \bibinfo {author} {\bibfnamefont {Y.}~\bibnamefont
  {Mudryk}},\ and\ \bibinfo {author} {\bibfnamefont {D.}~\bibnamefont
  {Paudyal}},\ }\bibfield  {title} {\bibinfo {title} {Making the most of the
  magnetic and lattice entropy changes},\ }\href@noop {} {\bibfield  {journal}
  {\bibinfo  {journal} {J. Magn. Magn. Mater.}\ }\textbf {\bibinfo {volume}
  {321}},\ \bibinfo {pages} {3541} (\bibinfo {year} {2009})}\BibitemShut
  {NoStop}%
\bibitem [{\citenamefont {Aliev}\ \emph {et~al.}(2021)\citenamefont {Aliev},
  \citenamefont {Khanov}, \citenamefont {Gamzatov}, \citenamefont {Batdalov},
  \citenamefont {Kurbanova}, \citenamefont {Yanushkevich},\ and\ \citenamefont
  {Govor}}]{aliev2021giant}%
  \BibitemOpen
  \bibfield  {author} {\bibinfo {author} {\bibfnamefont {A.}~\bibnamefont
  {Aliev}}, \bibinfo {author} {\bibfnamefont {L.}~\bibnamefont {Khanov}},
  \bibinfo {author} {\bibfnamefont {A.}~\bibnamefont {Gamzatov}}, \bibinfo
  {author} {\bibfnamefont {A.}~\bibnamefont {Batdalov}}, \bibinfo {author}
  {\bibfnamefont {D.}~\bibnamefont {Kurbanova}}, \bibinfo {author}
  {\bibfnamefont {K.}~\bibnamefont {Yanushkevich}},\ and\ \bibinfo {author}
  {\bibfnamefont {G.}~\bibnamefont {Govor}},\ }\bibfield  {title} {\bibinfo
  {title} {Giant magnetocaloric effect in {M}n{A}s$_{1-x}${P}$_x$ in a cyclic
  magnetic field: Lattice and magnetic contributions and degradation of the
  effect},\ }\href@noop {} {\bibfield  {journal} {\bibinfo  {journal} {Appl.
  Phys. Lett}\ }\textbf {\bibinfo {volume} {118}},\ \bibinfo {pages} {072404}
  (\bibinfo {year} {2021})}\BibitemShut {NoStop}%
\bibitem [{\citenamefont {Ma}\ \emph {et~al.}(2008)\citenamefont {Ma},
  \citenamefont {Woo},\ and\ \citenamefont {Dudarev}}]{ma2008large}%
  \BibitemOpen
  \bibfield  {author} {\bibinfo {author} {\bibfnamefont {P.-W.}\ \bibnamefont
  {Ma}}, \bibinfo {author} {\bibfnamefont {C.}~\bibnamefont {Woo}},\ and\
  \bibinfo {author} {\bibfnamefont {S.}~\bibnamefont {Dudarev}},\ }\bibfield
  {title} {\bibinfo {title} {Large-scale simulation of the spin-lattice
  dynamics in ferromagnetic iron},\ }\href@noop {} {\bibfield  {journal}
  {\bibinfo  {journal} {Physical Review B}\ }\textbf {\bibinfo {volume} {78}},\
  \bibinfo {pages} {024434} (\bibinfo {year} {2008})}\BibitemShut {NoStop}%
\bibitem [{\citenamefont {Wu}\ \emph {et~al.}(2018)\citenamefont {Wu},
  \citenamefont {Liu},\ and\ \citenamefont {Luo}}]{wu2018magnon}%
  \BibitemOpen
  \bibfield  {author} {\bibinfo {author} {\bibfnamefont {X.}~\bibnamefont
  {Wu}}, \bibinfo {author} {\bibfnamefont {Z.}~\bibnamefont {Liu}},\ and\
  \bibinfo {author} {\bibfnamefont {T.}~\bibnamefont {Luo}},\ }\bibfield
  {title} {\bibinfo {title} {Magnon and phonon dispersion, lifetime, and
  thermal conductivity of iron from spin-lattice dynamics simulations},\
  }\href@noop {} {\bibfield  {journal} {\bibinfo  {journal} {Journal of Applied
  Physics}\ }\textbf {\bibinfo {volume} {123}},\ \bibinfo {pages} {085109}
  (\bibinfo {year} {2018})}\BibitemShut {NoStop}%
\bibitem [{\citenamefont {Hellsvik}\ \emph {et~al.}(2019)\citenamefont
  {Hellsvik}, \citenamefont {Thonig}, \citenamefont {Modin}, \citenamefont
  {Iu{\c{s}}an}, \citenamefont {Bergman}, \citenamefont {Eriksson},
  \citenamefont {Bergqvist},\ and\ \citenamefont
  {Delin}}]{hellsvik2019general}%
  \BibitemOpen
  \bibfield  {author} {\bibinfo {author} {\bibfnamefont {J.}~\bibnamefont
  {Hellsvik}}, \bibinfo {author} {\bibfnamefont {D.}~\bibnamefont {Thonig}},
  \bibinfo {author} {\bibfnamefont {K.}~\bibnamefont {Modin}}, \bibinfo
  {author} {\bibfnamefont {D.}~\bibnamefont {Iu{\c{s}}an}}, \bibinfo {author}
  {\bibfnamefont {A.}~\bibnamefont {Bergman}}, \bibinfo {author} {\bibfnamefont
  {O.}~\bibnamefont {Eriksson}}, \bibinfo {author} {\bibfnamefont
  {L.}~\bibnamefont {Bergqvist}},\ and\ \bibinfo {author} {\bibfnamefont
  {A.}~\bibnamefont {Delin}},\ }\bibfield  {title} {\bibinfo {title} {General
  method for atomistic spin-lattice dynamics with first-principles accuracy},\
  }\href@noop {} {\bibfield  {journal} {\bibinfo  {journal} {Physical Review
  B}\ }\textbf {\bibinfo {volume} {99}},\ \bibinfo {pages} {104302} (\bibinfo
  {year} {2019})}\BibitemShut {NoStop}%
\bibitem [{\citenamefont {Perera}\ \emph {et~al.}(2017)\citenamefont {Perera},
  \citenamefont {Nicholson}, \citenamefont {Eisenbach}, \citenamefont
  {Stocks},\ and\ \citenamefont {Landau}}]{perera2017collective}%
  \BibitemOpen
  \bibfield  {author} {\bibinfo {author} {\bibfnamefont {D.}~\bibnamefont
  {Perera}}, \bibinfo {author} {\bibfnamefont {D.~M.}\ \bibnamefont
  {Nicholson}}, \bibinfo {author} {\bibfnamefont {M.}~\bibnamefont
  {Eisenbach}}, \bibinfo {author} {\bibfnamefont {G.~M.}\ \bibnamefont
  {Stocks}},\ and\ \bibinfo {author} {\bibfnamefont {D.~P.}\ \bibnamefont
  {Landau}},\ }\bibfield  {title} {\bibinfo {title} {Collective dynamics in
  atomistic models with coupled translational and spin degrees of freedom},\
  }\href@noop {} {\bibfield  {journal} {\bibinfo  {journal} {Physical Review
  B}\ }\textbf {\bibinfo {volume} {95}},\ \bibinfo {pages} {014431} (\bibinfo
  {year} {2017})}\BibitemShut {NoStop}%
\bibitem [{\citenamefont {Nolting}\ and\ \citenamefont
  {Ramakanth}(2009)}]{nolting2009quantum}%
  \BibitemOpen
  \bibfield  {author} {\bibinfo {author} {\bibfnamefont {W.}~\bibnamefont
  {Nolting}}\ and\ \bibinfo {author} {\bibfnamefont {A.}~\bibnamefont
  {Ramakanth}},\ }\href@noop {} {\emph {\bibinfo {title} {Quantum theory of
  magnetism}}}\ (\bibinfo  {publisher} {Springer Science \& Business Media},\
  \bibinfo {year} {2009})\BibitemShut {NoStop}%
\bibitem [{\citenamefont {Parkin}(1991)}]{parkin1991systematic}%
  \BibitemOpen
  \bibfield  {author} {\bibinfo {author} {\bibfnamefont {S.~S.}\ \bibnamefont
  {Parkin}},\ }\bibfield  {title} {\bibinfo {title} {Systematic variation of
  the strength and oscillation period of indirect magnetic exchange coupling
  through the 3$d$, 4$d$, and 5$d$ transition metals},\ }\href@noop {}
  {\bibfield  {journal} {\bibinfo  {journal} {Phys. Rev. Lett.}\ }\textbf
  {\bibinfo {volume} {67}},\ \bibinfo {pages} {3598} (\bibinfo {year}
  {1991})}\BibitemShut {NoStop}%
\bibitem [{\citenamefont {Ruderman}\ and\ \citenamefont
  {Kittel}(1954)}]{ruderman1954indirect}%
  \BibitemOpen
  \bibfield  {author} {\bibinfo {author} {\bibfnamefont {M.~A.}\ \bibnamefont
  {Ruderman}}\ and\ \bibinfo {author} {\bibfnamefont {C.}~\bibnamefont
  {Kittel}},\ }\bibfield  {title} {\bibinfo {title} {Indirect exchange coupling
  of nuclear magnetic moments by conduction electrons},\ }\href@noop {}
  {\bibfield  {journal} {\bibinfo  {journal} {Phys. Rev.}\ }\textbf {\bibinfo
  {volume} {96}},\ \bibinfo {pages} {99} (\bibinfo {year} {1954})}\BibitemShut
  {NoStop}%
\bibitem [{\citenamefont {Lindg{\aa}rd}\ \emph {et~al.}(1975)\citenamefont
  {Lindg{\aa}rd}, \citenamefont {Harmon},\ and\ \citenamefont
  {Freeman}}]{lindgaard1975theoretical}%
  \BibitemOpen
  \bibfield  {author} {\bibinfo {author} {\bibfnamefont {P.-A.}\ \bibnamefont
  {Lindg{\aa}rd}}, \bibinfo {author} {\bibfnamefont {B.}~\bibnamefont
  {Harmon}},\ and\ \bibinfo {author} {\bibfnamefont {A.}~\bibnamefont
  {Freeman}},\ }\bibfield  {title} {\bibinfo {title} {Theoretical magnon
  dispersion curves for {Gd}},\ }\href@noop {} {\bibfield  {journal} {\bibinfo
  {journal} {Physical Review Letters}\ }\textbf {\bibinfo {volume} {35}},\
  \bibinfo {pages} {383} (\bibinfo {year} {1975})}\BibitemShut {NoStop}%
\bibitem [{\citenamefont {Hindmarch}\ and\ \citenamefont
  {Hickey}(2003)}]{hindmarch2003direct}%
  \BibitemOpen
  \bibfield  {author} {\bibinfo {author} {\bibfnamefont {A.}~\bibnamefont
  {Hindmarch}}\ and\ \bibinfo {author} {\bibfnamefont {B.}~\bibnamefont
  {Hickey}},\ }\bibfield  {title} {\bibinfo {title} {Direct experimental
  evidence for the {Ruderman-Kittel-Kasuya-Yosida} interaction in rare-earth
  metals},\ }\href@noop {} {\bibfield  {journal} {\bibinfo  {journal} {Physical
  Review Letters}\ }\textbf {\bibinfo {volume} {91}},\ \bibinfo {pages}
  {116601} (\bibinfo {year} {2003})}\BibitemShut {NoStop}%
\bibitem [{\citenamefont {Scheie}\ \emph {et~al.}(2022)\citenamefont {Scheie},
  \citenamefont {Laurell}, \citenamefont {McClarty}, \citenamefont {Granroth},
  \citenamefont {Stone}, \citenamefont {Moessner},\ and\ \citenamefont
  {Nagler}}]{scheie2022spin}%
  \BibitemOpen
  \bibfield  {author} {\bibinfo {author} {\bibfnamefont {A.}~\bibnamefont
  {Scheie}}, \bibinfo {author} {\bibfnamefont {P.}~\bibnamefont {Laurell}},
  \bibinfo {author} {\bibfnamefont {P.~A.}\ \bibnamefont {McClarty}}, \bibinfo
  {author} {\bibfnamefont {G.~E.}\ \bibnamefont {Granroth}}, \bibinfo {author}
  {\bibfnamefont {M.~B.}\ \bibnamefont {Stone}}, \bibinfo {author}
  {\bibfnamefont {R.}~\bibnamefont {Moessner}},\ and\ \bibinfo {author}
  {\bibfnamefont {S.~E.}\ \bibnamefont {Nagler}},\ }\bibfield  {title}
  {\bibinfo {title} {Spin-exchange {H}amiltonian and topological degeneracies
  in elemental gadolinium},\ }\href@noop {} {\bibfield  {journal} {\bibinfo
  {journal} {Physical Review B}\ }\textbf {\bibinfo {volume} {105}},\ \bibinfo
  {pages} {104402} (\bibinfo {year} {2022})}\BibitemShut {NoStop}%
\bibitem [{\citenamefont {Parkin}\ \emph {et~al.}(1990)\citenamefont {Parkin},
  \citenamefont {More},\ and\ \citenamefont {Roche}}]{parkin1990oscillations}%
  \BibitemOpen
  \bibfield  {author} {\bibinfo {author} {\bibfnamefont {S.}~\bibnamefont
  {Parkin}}, \bibinfo {author} {\bibfnamefont {N.}~\bibnamefont {More}},\ and\
  \bibinfo {author} {\bibfnamefont {K.}~\bibnamefont {Roche}},\ }\bibfield
  {title} {\bibinfo {title} {Oscillations in exchange coupling and
  magnetoresistance in metallic superlattice structures: {C}o/{R}u, {C}o/{C}r,
  and {F}e/{C}r},\ }\href@noop {} {\bibfield  {journal} {\bibinfo  {journal}
  {Phys. Rev. Lett.}\ }\textbf {\bibinfo {volume} {64}},\ \bibinfo {pages}
  {2304} (\bibinfo {year} {1990})}\BibitemShut {NoStop}%
\bibitem [{\citenamefont {Martinho~Vieira}\ \emph {et~al.}(2022)\citenamefont
  {Martinho~Vieira}, \citenamefont {Eriksson}, \citenamefont {Bj{\"o}rkman},
  \citenamefont {Bergman},\ and\ \citenamefont
  {Herper}}]{martinho2022realistic}%
  \BibitemOpen
  \bibfield  {author} {\bibinfo {author} {\bibfnamefont {R.}~\bibnamefont
  {Martinho~Vieira}}, \bibinfo {author} {\bibfnamefont {O.}~\bibnamefont
  {Eriksson}}, \bibinfo {author} {\bibfnamefont {T.}~\bibnamefont
  {Bj{\"o}rkman}}, \bibinfo {author} {\bibfnamefont {A.}~\bibnamefont
  {Bergman}},\ and\ \bibinfo {author} {\bibfnamefont {H.~C.}\ \bibnamefont
  {Herper}},\ }\bibfield  {title} {\bibinfo {title} {Realistic first-principles
  calculations of the magnetocaloric effect: applications to hcp {G}d},\
  }\href@noop {} {\bibfield  {journal} {\bibinfo  {journal} {Mater. Res.
  Lett.}\ }\textbf {\bibinfo {volume} {10}},\ \bibinfo {pages} {156} (\bibinfo
  {year} {2022})}\BibitemShut {NoStop}%
\bibitem [{\citenamefont {Kresse}\ and\ \citenamefont
  {Furthm{\"u}ller}(1996)}]{kresse1996efficient}%
  \BibitemOpen
  \bibfield  {author} {\bibinfo {author} {\bibfnamefont {G.}~\bibnamefont
  {Kresse}}\ and\ \bibinfo {author} {\bibfnamefont {J.}~\bibnamefont
  {Furthm{\"u}ller}},\ }\bibfield  {title} {\bibinfo {title} {Efficient
  iterative schemes for ab-initio total-energy calculations using a plane-wave
  basis set},\ }\href@noop {} {\bibfield  {journal} {\bibinfo  {journal} {Phys.
  Rev. B}\ }\textbf {\bibinfo {volume} {54}},\ \bibinfo {pages} {11169}
  (\bibinfo {year} {1996})}\BibitemShut {NoStop}%
\bibitem [{\citenamefont {Bl{\"o}chl}(1994)}]{blochl1994projector}%
  \BibitemOpen
  \bibfield  {author} {\bibinfo {author} {\bibfnamefont {P.~E.}\ \bibnamefont
  {Bl{\"o}chl}},\ }\bibfield  {title} {\bibinfo {title} {Projector
  augmented-wave method},\ }\href@noop {} {\bibfield  {journal} {\bibinfo
  {journal} {Phys. Rev. B}\ }\textbf {\bibinfo {volume} {50}},\ \bibinfo
  {pages} {17953} (\bibinfo {year} {1994})}\BibitemShut {NoStop}%
\bibitem [{\citenamefont {Pack}\ and\ \citenamefont
  {Monkhorst}(1977)}]{pack1977special}%
  \BibitemOpen
  \bibfield  {author} {\bibinfo {author} {\bibfnamefont {J.~D.}\ \bibnamefont
  {Pack}}\ and\ \bibinfo {author} {\bibfnamefont {H.~J.}\ \bibnamefont
  {Monkhorst}},\ }\bibfield  {title} {\bibinfo {title} {``special points for
  {B}rillouin-zone integrations"—a reply},\ }\href@noop {} {\bibfield
  {journal} {\bibinfo  {journal} {Phys. Rev. B}\ }\textbf {\bibinfo {volume}
  {16}},\ \bibinfo {pages} {1748} (\bibinfo {year} {1977})}\BibitemShut
  {NoStop}%
\bibitem [{\citenamefont {Ebert}\ \emph {et~al.}(2011)\citenamefont {Ebert},
  \citenamefont {Koedderitzsch},\ and\ \citenamefont
  {Minar}}]{ebert2011calculating}%
  \BibitemOpen
  \bibfield  {author} {\bibinfo {author} {\bibfnamefont {H.}~\bibnamefont
  {Ebert}}, \bibinfo {author} {\bibfnamefont {D.}~\bibnamefont
  {Koedderitzsch}},\ and\ \bibinfo {author} {\bibfnamefont {J.}~\bibnamefont
  {Minar}},\ }\bibfield  {title} {\bibinfo {title} {Calculating condensed
  matter properties using the {KKR}-{G}reen's function method—recent
  developments and applications},\ }\href@noop {} {\bibfield  {journal}
  {\bibinfo  {journal} {Rep. Prog. Phys.}\ }\textbf {\bibinfo {volume} {74}},\
  \bibinfo {pages} {096501} (\bibinfo {year} {2011})}\BibitemShut {NoStop}%
\bibitem [{\citenamefont {Plimpton}(1995)}]{plimpton1995fast}%
  \BibitemOpen
  \bibfield  {author} {\bibinfo {author} {\bibfnamefont {S.}~\bibnamefont
  {Plimpton}},\ }\bibfield  {title} {\bibinfo {title} {Fast parallel algorithms
  for short-range molecular dynamics},\ }\href@noop {} {\bibfield  {journal}
  {\bibinfo  {journal} {J. Comput. Phys.}\ }\textbf {\bibinfo {volume} {117}},\
  \bibinfo {pages} {1} (\bibinfo {year} {1995})}\BibitemShut {NoStop}%
\bibitem [{\citenamefont {Tranchida}\ \emph {et~al.}(2018)\citenamefont
  {Tranchida}, \citenamefont {Plimpton}, \citenamefont {Thibaudeau},\ and\
  \citenamefont {Thompson}}]{tranchida2018massively}%
  \BibitemOpen
  \bibfield  {author} {\bibinfo {author} {\bibfnamefont {J.}~\bibnamefont
  {Tranchida}}, \bibinfo {author} {\bibfnamefont {S.~J.}\ \bibnamefont
  {Plimpton}}, \bibinfo {author} {\bibfnamefont {P.}~\bibnamefont
  {Thibaudeau}},\ and\ \bibinfo {author} {\bibfnamefont {A.~P.}\ \bibnamefont
  {Thompson}},\ }\bibfield  {title} {\bibinfo {title} {Massively parallel
  symplectic algorithm for coupled magnetic spin dynamics and molecular
  dynamics},\ }\href@noop {} {\bibfield  {journal} {\bibinfo  {journal}
  {Journal of Computational Physics}\ }\textbf {\bibinfo {volume} {372}},\
  \bibinfo {pages} {406} (\bibinfo {year} {2018})}\BibitemShut {NoStop}%
\bibitem [{\citenamefont {Wang}\ \emph {et~al.}(2010)\citenamefont {Wang},
  \citenamefont {Ma},\ and\ \citenamefont {Woo}}]{wang2010exchange}%
  \BibitemOpen
  \bibfield  {author} {\bibinfo {author} {\bibfnamefont {H.}~\bibnamefont
  {Wang}}, \bibinfo {author} {\bibfnamefont {P.-W.}\ \bibnamefont {Ma}},\ and\
  \bibinfo {author} {\bibfnamefont {C.}~\bibnamefont {Woo}},\ }\bibfield
  {title} {\bibinfo {title} {Exchange interaction function for spin-lattice
  coupling in {BCC} iron},\ }\href@noop {} {\bibfield  {journal} {\bibinfo
  {journal} {Phys. Rev. B}\ }\textbf {\bibinfo {volume} {82}},\ \bibinfo
  {pages} {144304} (\bibinfo {year} {2010})}\BibitemShut {NoStop}%
\bibitem [{\citenamefont {Pajda}\ \emph {et~al.}(2001)\citenamefont {Pajda},
  \citenamefont {Kudrnovsk{\`y}}, \citenamefont {Turek}, \citenamefont
  {Drchal},\ and\ \citenamefont {Bruno}}]{pajda2001ab}%
  \BibitemOpen
  \bibfield  {author} {\bibinfo {author} {\bibfnamefont {M.}~\bibnamefont
  {Pajda}}, \bibinfo {author} {\bibfnamefont {J.}~\bibnamefont
  {Kudrnovsk{\`y}}}, \bibinfo {author} {\bibfnamefont {I.}~\bibnamefont
  {Turek}}, \bibinfo {author} {\bibfnamefont {V.}~\bibnamefont {Drchal}},\ and\
  \bibinfo {author} {\bibfnamefont {P.}~\bibnamefont {Bruno}},\ }\bibfield
  {title} {\bibinfo {title} {Ab initio calculations of exchange interactions,
  spin-wave stiffness constants, and {C}urie temperatures of {F}e, {Co}, and
  {Ni}},\ }\href@noop {} {\bibfield  {journal} {\bibinfo  {journal} {Phys. Rev.
  B}\ }\textbf {\bibinfo {volume} {64}},\ \bibinfo {pages} {174402} (\bibinfo
  {year} {2001})}\BibitemShut {NoStop}%
\bibitem [{\citenamefont {Mryasov}\ \emph {et~al.}(1996)\citenamefont
  {Mryasov}, \citenamefont {Freeman},\ and\ \citenamefont
  {Liechtenstein}}]{mryasov1996theory}%
  \BibitemOpen
  \bibfield  {author} {\bibinfo {author} {\bibfnamefont {O.}~\bibnamefont
  {Mryasov}}, \bibinfo {author} {\bibfnamefont {A.~J.}\ \bibnamefont
  {Freeman}},\ and\ \bibinfo {author} {\bibfnamefont {A.}~\bibnamefont
  {Liechtenstein}},\ }\bibfield  {title} {\bibinfo {title} {Theory of
  non-{H}eisenberg exchange: Results for localized and itinerant magnets},\
  }\href@noop {} {\bibfield  {journal} {\bibinfo  {journal} {J. Appl. Phys.}\
  }\textbf {\bibinfo {volume} {79}},\ \bibinfo {pages} {4805} (\bibinfo {year}
  {1996})}\BibitemShut {NoStop}%
\bibitem [{\citenamefont {Frota-Pess{\^o}a}\ \emph {et~al.}(2000)\citenamefont
  {Frota-Pess{\^o}a}, \citenamefont {Muniz},\ and\ \citenamefont
  {Kudrnovsk{\`y}}}]{frota2000exchange}%
  \BibitemOpen
  \bibfield  {author} {\bibinfo {author} {\bibfnamefont {S.}~\bibnamefont
  {Frota-Pess{\^o}a}}, \bibinfo {author} {\bibfnamefont {R.}~\bibnamefont
  {Muniz}},\ and\ \bibinfo {author} {\bibfnamefont {J.}~\bibnamefont
  {Kudrnovsk{\`y}}},\ }\bibfield  {title} {\bibinfo {title} {Exchange coupling
  in transition-metal ferromagnets},\ }\href@noop {} {\bibfield  {journal}
  {\bibinfo  {journal} {Phys. Rev. B}\ }\textbf {\bibinfo {volume} {62}},\
  \bibinfo {pages} {5293} (\bibinfo {year} {2000})}\BibitemShut {NoStop}%
\bibitem [{\citenamefont {Kvashnin}\ \emph {et~al.}(2016)\citenamefont
  {Kvashnin}, \citenamefont {Cardias}, \citenamefont {Szilva}, \citenamefont
  {Di~Marco}, \citenamefont {Katsnelson}, \citenamefont {Lichtenstein},
  \citenamefont {Nordstr{\"o}m}, \citenamefont {Klautau},\ and\ \citenamefont
  {Eriksson}}]{kvashnin2016microscopic}%
  \BibitemOpen
  \bibfield  {author} {\bibinfo {author} {\bibfnamefont {Y.~O.}\ \bibnamefont
  {Kvashnin}}, \bibinfo {author} {\bibfnamefont {R.}~\bibnamefont {Cardias}},
  \bibinfo {author} {\bibfnamefont {A.}~\bibnamefont {Szilva}}, \bibinfo
  {author} {\bibfnamefont {I.}~\bibnamefont {Di~Marco}}, \bibinfo {author}
  {\bibfnamefont {M.}~\bibnamefont {Katsnelson}}, \bibinfo {author}
  {\bibfnamefont {A.}~\bibnamefont {Lichtenstein}}, \bibinfo {author}
  {\bibfnamefont {L.}~\bibnamefont {Nordstr{\"o}m}}, \bibinfo {author}
  {\bibfnamefont {A.}~\bibnamefont {Klautau}},\ and\ \bibinfo {author}
  {\bibfnamefont {O.}~\bibnamefont {Eriksson}},\ }\bibfield  {title} {\bibinfo
  {title} {Microscopic origin of {H}eisenberg and non-{H}eisenberg exchange
  interactions in ferromagnetic bcc {F}e},\ }\href@noop {} {\bibfield
  {journal} {\bibinfo  {journal} {Physical Review Letters}\ }\textbf {\bibinfo
  {volume} {116}},\ \bibinfo {pages} {217202} (\bibinfo {year}
  {2016})}\BibitemShut {NoStop}%
\bibitem [{\citenamefont {Kasuya}(1956)}]{kasuya1956theory}%
  \BibitemOpen
  \bibfield  {author} {\bibinfo {author} {\bibfnamefont {T.}~\bibnamefont
  {Kasuya}},\ }\bibfield  {title} {\bibinfo {title} {A theory of metallic
  ferro-and antiferromagnetism on {Z}ener's model},\ }\href@noop {} {\bibfield
  {journal} {\bibinfo  {journal} {Prog. Theor. Phys.}\ }\textbf {\bibinfo
  {volume} {16}},\ \bibinfo {pages} {45} (\bibinfo {year} {1956})}\BibitemShut
  {NoStop}%
\bibitem [{\citenamefont {Yosida}(1957)}]{yosida1957magnetic}%
  \BibitemOpen
  \bibfield  {author} {\bibinfo {author} {\bibfnamefont {K.}~\bibnamefont
  {Yosida}},\ }\bibfield  {title} {\bibinfo {title} {Magnetic properties of
  {C}u-{M}n alloys},\ }\href@noop {} {\bibfield  {journal} {\bibinfo  {journal}
  {Phys. Rev.}\ }\textbf {\bibinfo {volume} {106}},\ \bibinfo {pages} {893}
  (\bibinfo {year} {1957})}\BibitemShut {NoStop}%
\bibitem [{\citenamefont {Kurz}\ \emph {et~al.}(2002)\citenamefont {Kurz},
  \citenamefont {Bihlmayer},\ and\ \citenamefont
  {Bl{\"u}gel}}]{kurz2002magnetism}%
  \BibitemOpen
  \bibfield  {author} {\bibinfo {author} {\bibfnamefont {P.}~\bibnamefont
  {Kurz}}, \bibinfo {author} {\bibfnamefont {G.}~\bibnamefont {Bihlmayer}},\
  and\ \bibinfo {author} {\bibfnamefont {S.}~\bibnamefont {Bl{\"u}gel}},\
  }\bibfield  {title} {\bibinfo {title} {Magnetism and electronic structure of
  hcp {G}d and the {G}d (0001) surface},\ }\href@noop {} {\bibfield  {journal}
  {\bibinfo  {journal} {J. Phys.: Condens. Matter}\ }\textbf {\bibinfo {volume}
  {14}},\ \bibinfo {pages} {6353} (\bibinfo {year} {2002})}\BibitemShut
  {NoStop}%
\bibitem [{\citenamefont {Zhang}\ \emph {et~al.}(2017)\citenamefont {Zhang},
  \citenamefont {Jenkins}, \citenamefont {Bennett},\ and\ \citenamefont
  {Bai}}]{zhang2017manifestation}%
  \BibitemOpen
  \bibfield  {author} {\bibinfo {author} {\bibfnamefont {G.}~\bibnamefont
  {Zhang}}, \bibinfo {author} {\bibfnamefont {T.}~\bibnamefont {Jenkins}},
  \bibinfo {author} {\bibfnamefont {M.}~\bibnamefont {Bennett}},\ and\ \bibinfo
  {author} {\bibfnamefont {Y.}~\bibnamefont {Bai}},\ }\bibfield  {title}
  {\bibinfo {title} {Manifestation of intra-atomic 5d6s-4f exchange coupling in
  photoexcited gadolinium},\ }\href@noop {} {\bibfield  {journal} {\bibinfo
  {journal} {J. Phys.: Condens. Matter}\ }\textbf {\bibinfo {volume} {29}},\
  \bibinfo {pages} {495807} (\bibinfo {year} {2017})}\BibitemShut {NoStop}%
\bibitem [{\citenamefont {Turek}\ \emph {et~al.}(2006)\citenamefont {Turek},
  \citenamefont {Kudrnovsk{\`y}}, \citenamefont {Drchal},\ and\ \citenamefont
  {Bruno}}]{turek2006exchange}%
  \BibitemOpen
  \bibfield  {author} {\bibinfo {author} {\bibfnamefont {I.}~\bibnamefont
  {Turek}}, \bibinfo {author} {\bibfnamefont {J.}~\bibnamefont
  {Kudrnovsk{\`y}}}, \bibinfo {author} {\bibfnamefont {V.}~\bibnamefont
  {Drchal}},\ and\ \bibinfo {author} {\bibfnamefont {P.}~\bibnamefont
  {Bruno}},\ }\bibfield  {title} {\bibinfo {title} {Exchange interactions, spin
  waves, and transition temperatures in itinerant magnets},\ }\href@noop {}
  {\bibfield  {journal} {\bibinfo  {journal} {Philos. Mag.}\ }\textbf {\bibinfo
  {volume} {86}},\ \bibinfo {pages} {1713} (\bibinfo {year}
  {2006})}\BibitemShut {NoStop}%
\bibitem [{\citenamefont {Kvashnin}\ \emph {et~al.}(2015)\citenamefont
  {Kvashnin}, \citenamefont {Gr{\aa}n{\"a}s}, \citenamefont {Di~Marco},
  \citenamefont {Katsnelson}, \citenamefont {Lichtenstein},\ and\ \citenamefont
  {Eriksson}}]{kvashnin2015exchange}%
  \BibitemOpen
  \bibfield  {author} {\bibinfo {author} {\bibfnamefont {Y.~O.}\ \bibnamefont
  {Kvashnin}}, \bibinfo {author} {\bibfnamefont {O.}~\bibnamefont
  {Gr{\aa}n{\"a}s}}, \bibinfo {author} {\bibfnamefont {I.}~\bibnamefont
  {Di~Marco}}, \bibinfo {author} {\bibfnamefont {M.}~\bibnamefont
  {Katsnelson}}, \bibinfo {author} {\bibfnamefont {A.}~\bibnamefont
  {Lichtenstein}},\ and\ \bibinfo {author} {\bibfnamefont {O.}~\bibnamefont
  {Eriksson}},\ }\bibfield  {title} {\bibinfo {title} {Exchange parameters of
  strongly correlated materials: Extraction from spin-polarized density
  functional theory plus dynamical mean-field theory},\ }\href@noop {}
  {\bibfield  {journal} {\bibinfo  {journal} {Phys.Rev. B}\ }\textbf {\bibinfo
  {volume} {91}},\ \bibinfo {pages} {125133} (\bibinfo {year}
  {2015})}\BibitemShut {NoStop}%
\bibitem [{\citenamefont {Gottschall}\ \emph {et~al.}(2019)\citenamefont
  {Gottschall}, \citenamefont {Skokov}, \citenamefont {Fries}, \citenamefont
  {Taubel}, \citenamefont {Radulov}, \citenamefont {Scheibel}, \citenamefont
  {Benke}, \citenamefont {Riegg},\ and\ \citenamefont
  {Gutfleisch}}]{gottschall2019making}%
  \BibitemOpen
  \bibfield  {author} {\bibinfo {author} {\bibfnamefont {T.}~\bibnamefont
  {Gottschall}}, \bibinfo {author} {\bibfnamefont {K.~P.}\ \bibnamefont
  {Skokov}}, \bibinfo {author} {\bibfnamefont {M.}~\bibnamefont {Fries}},
  \bibinfo {author} {\bibfnamefont {A.}~\bibnamefont {Taubel}}, \bibinfo
  {author} {\bibfnamefont {I.}~\bibnamefont {Radulov}}, \bibinfo {author}
  {\bibfnamefont {F.}~\bibnamefont {Scheibel}}, \bibinfo {author}
  {\bibfnamefont {D.}~\bibnamefont {Benke}}, \bibinfo {author} {\bibfnamefont
  {S.}~\bibnamefont {Riegg}},\ and\ \bibinfo {author} {\bibfnamefont
  {O.}~\bibnamefont {Gutfleisch}},\ }\bibfield  {title} {\bibinfo {title}
  {Making a cool choice: the materials library of magnetic refrigeration},\
  }\href@noop {} {\bibfield  {journal} {\bibinfo  {journal} {Adv. Energy
  Mater.}\ }\textbf {\bibinfo {volume} {9}},\ \bibinfo {pages} {1901322}
  (\bibinfo {year} {2019})}\BibitemShut {NoStop}%
\bibitem [{\citenamefont {Baskes}\ and\ \citenamefont
  {Johnson}(1994)}]{baskes1994modified}%
  \BibitemOpen
  \bibfield  {author} {\bibinfo {author} {\bibfnamefont {M.}~\bibnamefont
  {Baskes}}\ and\ \bibinfo {author} {\bibfnamefont {R.}~\bibnamefont
  {Johnson}},\ }\bibfield  {title} {\bibinfo {title} {Modified embedded atom
  potentials for {HCP} metals},\ }\href@noop {} {\bibfield  {journal} {\bibinfo
   {journal} {Model. Simul. Mat. Sci. Eng.}\ }\textbf {\bibinfo {volume} {2}},\
  \bibinfo {pages} {147} (\bibinfo {year} {1994})}\BibitemShut {NoStop}%
\bibitem [{\citenamefont {Kong}(2011)}]{kong2011phonon}%
  \BibitemOpen
  \bibfield  {author} {\bibinfo {author} {\bibfnamefont {L.~T.}\ \bibnamefont
  {Kong}},\ }\bibfield  {title} {\bibinfo {title} {Phonon dispersion measured
  directly from molecular dynamics simulations},\ }\href@noop {} {\bibfield
  {journal} {\bibinfo  {journal} {Computer Physics Communications}\ }\textbf
  {\bibinfo {volume} {182}},\ \bibinfo {pages} {2201} (\bibinfo {year}
  {2011})}\BibitemShut {NoStop}%
\bibitem [{\citenamefont {Wallace}(2002)}]{wallace2002statistical}%
  \BibitemOpen
  \bibfield  {author} {\bibinfo {author} {\bibfnamefont {D.~C.}\ \bibnamefont
  {Wallace}},\ }\href@noop {} {\emph {\bibinfo {title} {Statistical physics of
  crystals and liquids: a guide to highly accurate equations of state}}}\
  (\bibinfo  {publisher} {World Scientific},\ \bibinfo {year}
  {2002})\BibitemShut {NoStop}%
\bibitem [{\citenamefont {Hui}\ and\ \citenamefont
  {Allen}(1975)}]{hui1975thermodynamics}%
  \BibitemOpen
  \bibfield  {author} {\bibinfo {author} {\bibfnamefont {J.}~\bibnamefont
  {Hui}}\ and\ \bibinfo {author} {\bibfnamefont {P.~B.}\ \bibnamefont
  {Allen}},\ }\bibfield  {title} {\bibinfo {title} {Thermodynamics of
  anharmonic crystals with application to {N}b},\ }\href@noop {} {\bibfield
  {journal} {\bibinfo  {journal} {Journal of Physics C: Solid State Physics}\
  }\textbf {\bibinfo {volume} {8}},\ \bibinfo {pages} {2923} (\bibinfo {year}
  {1975})}\BibitemShut {NoStop}%
\end{thebibliography}%

\end{document}



\title{Supplementary Information: Indirect Exchange Interaction Leads to Large Lattice Contribution to Magnetocaloric Entropy Change} 

\author{Lokanath Patra}
\affiliation{Department of Mechanical Engineering, University of California, Santa Barbara, CA 93106, USA}

\author{Bolin Liao}
\email{bliao@ucsb.edu} \affiliation{Department of Mechanical Engineering, University of California, Santa Barbara, CA 93106, USA}

\maketitle



\section{Computational Methods}

\subsection{Spin Lattice Dynamics Simulations}
In this paper, we performed SLD simulations, where the spin degrees of freedom are coupled to the lattice degrees of freedom.~\cite{antropov1995ab,ma2008large} For this purpose, we ran our simulations using the SPIN package recently added to the software LAMMPS.~\cite{tranchida2018massively} Under this framework, one can introduce magnetic effects in a classical molecular dynamics simulation through a generalized Hamiltonian:
\begin{equation}
    \mathcal{H} = \sum_{i=1}^{N}\frac{|\textbf{\textit{p}}_i|^2}{2m_i} + \sum_{i, j, i\neq 1}^{N} V (r_{ij}) + \mathcal{H}_{mag}
\end{equation}

The first term is the kinetic energy of the atoms and the second term is a classical interatomic potential describing the mechanical interactions between the atoms. The last term is a magnetic Hamiltonian, which can contain several terms,
accounting for spin-spin exchange interactions, magnetic anisotropy, Zeeman, dipolar, Dzyaloshinskii-Moriya, and magnetoelectric interactions. In the present work, we have only considered the spin-spin exchange interactions and the interaction with external magnetic fields. The simplified magnetic Hamiltonian used in the present work is given by

\begin{equation}
    \mathcal{H}_{mag} = -\sum_{i, j, i\neq 1}^{N} \textbf{\textit{J}}(r_{ij})\textbf{\textit{S}}_i.\textbf{\textit{S}}_j + \mathcal{H}_{ext}.
\end{equation}

The first term is a Heisenberg Hamiltonian accounting for
spin-spin interactions, where \textbf{\textit{S}}$_i$ (\textbf{\textit{S}}$_j$) is the normalized spin vector
of spin $i$ ($j$) and $J(r_{ij})$ is the Heisenberg magnetic coupling exchange constant, which depends on the distance $r_{ij}$ between atoms $i$ and $j$. The second term accounts for the effect of the external magnetic field.

The central aspect of this simulation scheme is the addition of a classical spin vector $\textbf{\textit{s}}_i$ to each atom $i$. This enables magnetic degrees of freedom to be explicitly treated and added to the atomic degrees of freedom, momentum $\mathbf{p}_i$, and position $\textbf{\textit{r}}_i$. The equations of motion (EOM) can be derived from the Hamiltonian of Eq. (1):
\begin{equation}
    \frac{dr_i}{dt} = \frac{p_i}{m_i},
\end{equation}
\begin{equation}
    \frac{dp_i}{dt} = \sum_{i, j, i\neq 1}^{N}\left [\frac{dV (r_{ij})}{dr_{ij}} + \frac{dJ (r_{ij})}{dr_{ij}}\textbf{\textit{s}}_i.\textbf{\textit{s}}_j \right] \textbf{\textit{e}}_i,
\end{equation}
\begin{equation}
    \frac{d\textbf{\textit{s}}_i}{dt} = \boldsymbol{\omega}_i \times \textbf{\textit{s}}_i,
\end{equation}
where \textbf{\textit{e}}$_i$ is a unit vector along the direction of the vector \textbf{\textit{r}}$_{ij}$
and $\omega_i$ is the analog of a spin force applied to spin i, defined as
\begin{equation}
   \boldsymbol{\omega}_i = -\frac{1}{\hbar} \frac{\partial\mathcal{H}_{mag}}{\partial\boldsymbol{s}_i}. 
\end{equation}
We set up a cluster of bcc Fe and hcp Gd crystals with a radius of 3.5~\AA. The crystals were then relaxed for 50 ps at a temperature of 300 K and a pressure of 0 bar with a time step of 1 fs. We used a Langevin thermostat to control the temperature and a Berendsen barostat to keep the crystal at zero pressure. For the temperature dependence analysis, the crystals are heated to the desired temperature and then equilibrated for 10 ps, with a barostat at zero pressure. We determined local magnetic moments according to the approach presented in our previous study. All SLD simulations were performed using the SPIN package of the LAMMPS simulation tool.

\subsection{SLD Simulations of bcc Fe and hcp Gd}
The magnetization ($M$) versus temperature ($T$) curves for bcc Fe are shown in Fig.~S1 with different applied external fields. From this data, the entropy changes were calculated as a function of the applied field (Fig.~S2) using the Maxwell's equation (Eqn.~1 in the main text). An entropy change of $\sim$ -0.3\,J\,kg$^{-1}$\,K$^{-1}$ was calculated for bcc Fe. Similar calculations were also performed for hcp Gd resulting in M versus T curves (Fig.~S3) and $\Delta S_T$ versus T curves (Fig.~S4) as functions of externally applied magnetic fields. A $\Delta S_T$ of $\sim$ 10\,J\,kg$^{-1}$\,K$^{-1}$ was calculated for a field change of 5\,T for hcp Gd.

\begin{figure}[!htb]
\includegraphics[scale=0.5]{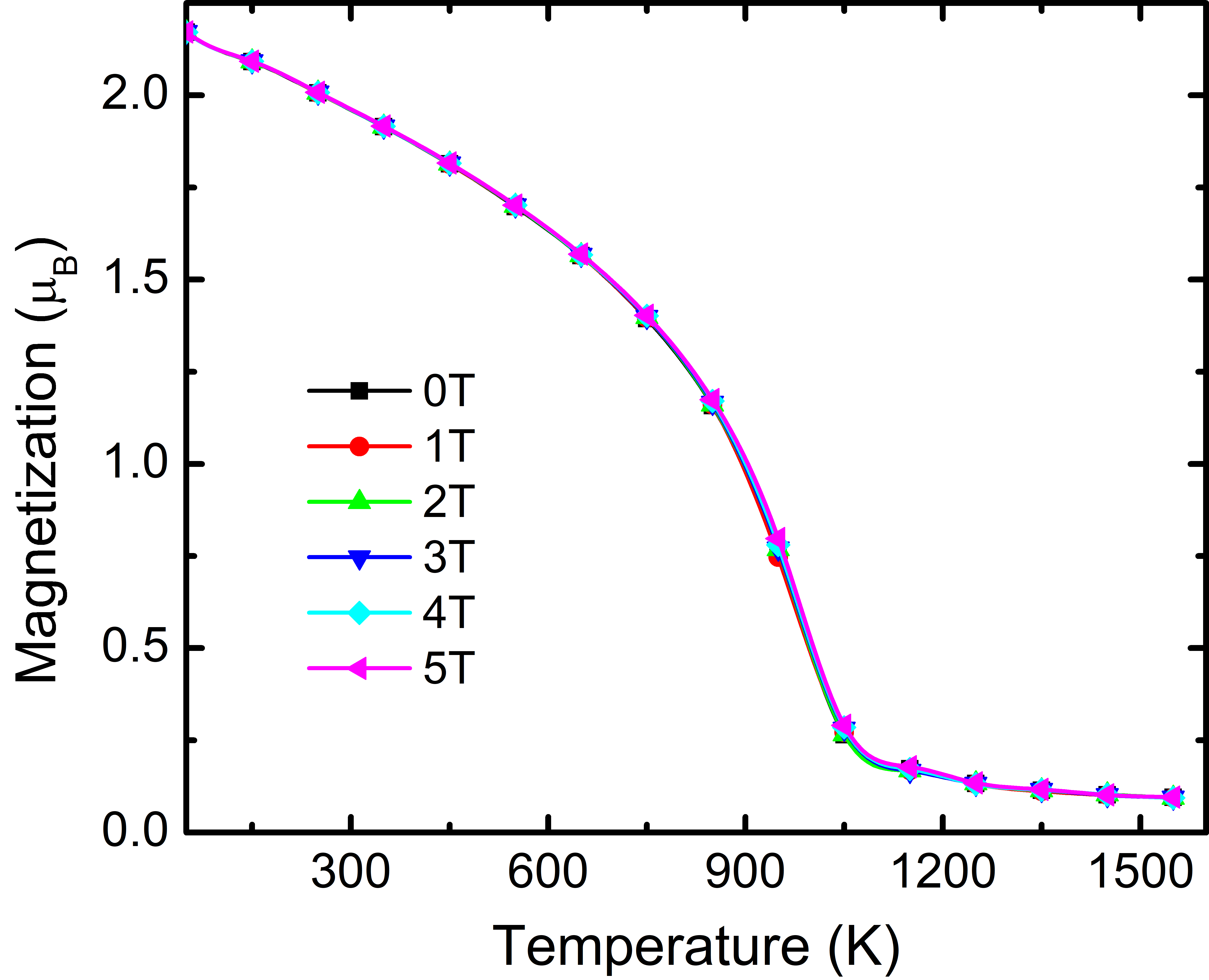}
\caption{The magnetic-field-dependent magnetization versus temperature curves for bcc Fe simulated with the SLD approach. From these curves, the magnetocaloric entropy change can be calculated.} 
\label{fig:figS1}
\end{figure}

\begin{figure}[!htb]
\includegraphics[scale=0.5]{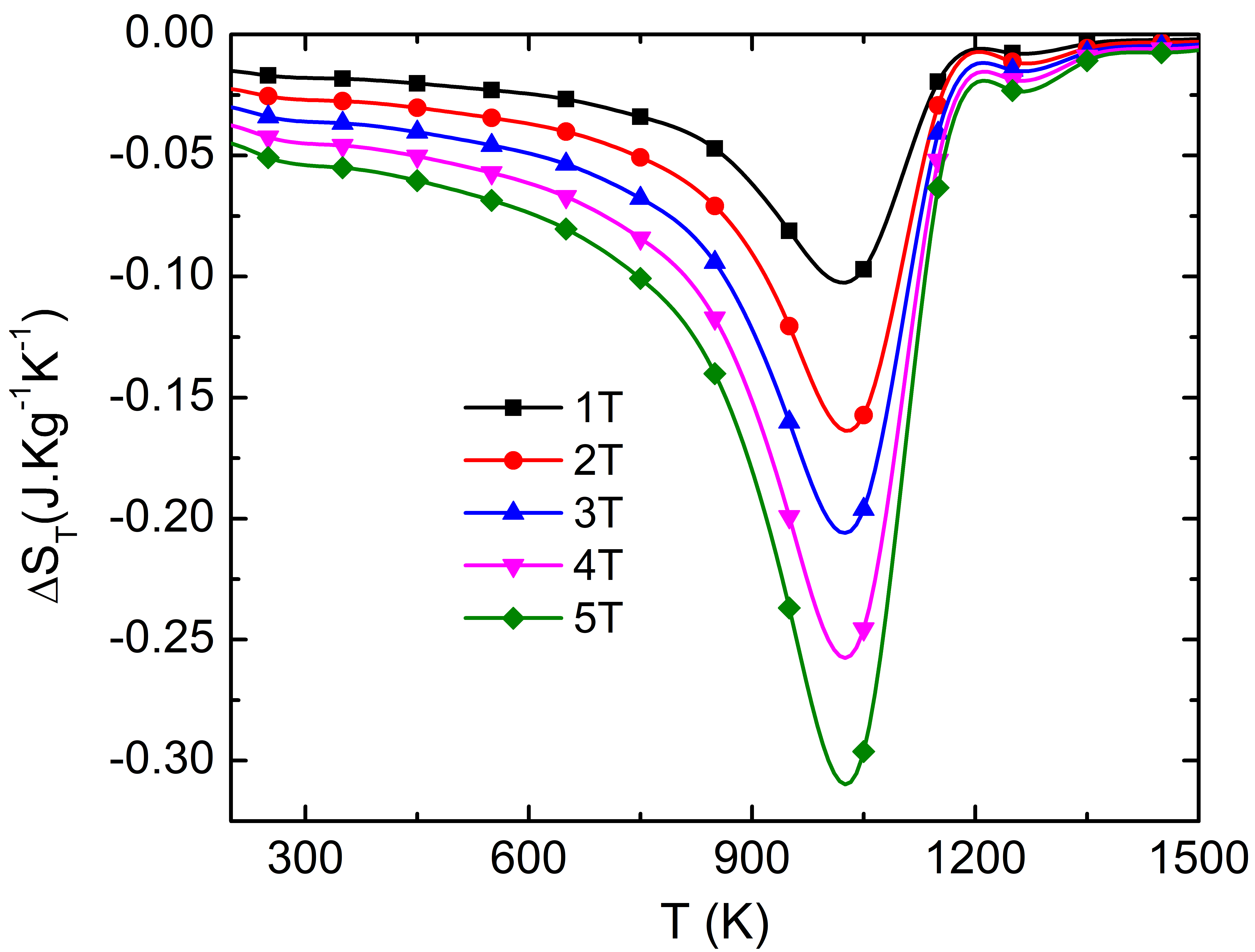}
\caption{The total magnetocaloric entropy change as a function of applied external magnetic fields for bcc Fe simulated with the SLD approach.} 
\label{fig:figS2}
\end{figure}

\begin{figure}[!htb]
\includegraphics[scale=0.5]{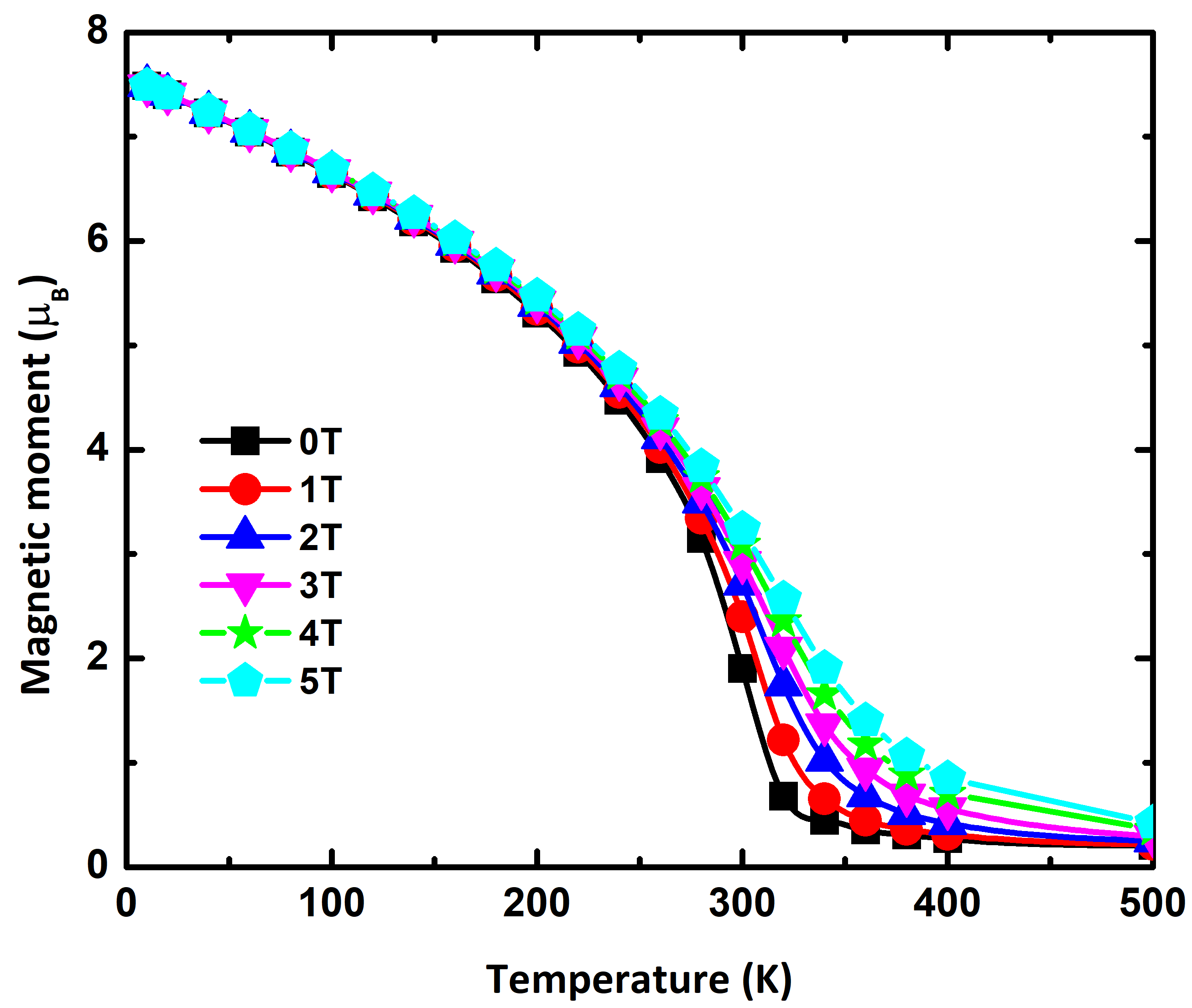}
\caption{The magnetic-field-dependent magnetization versus temperature curves for hcp Gd simulated with the SLD approach. } 
\label{fig:figS3}
\end{figure}

\begin{figure}[!htb]
\includegraphics[scale=0.5]{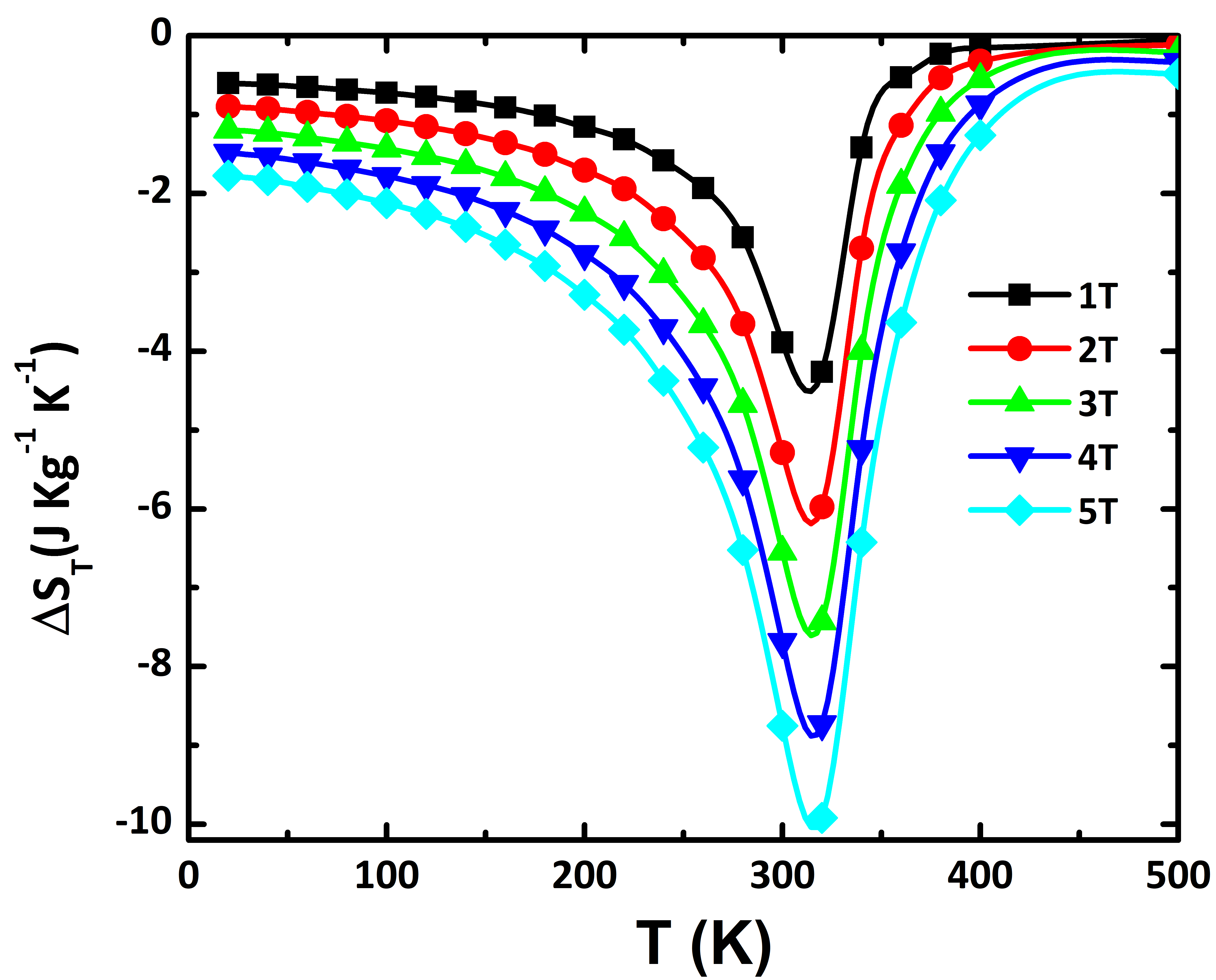}
\caption{The total magnetocaloric entropy change as a function of applied external magnetic fields for hcp Gd simulated with the SLD approach.} 
\label{fig:figS4}
\end{figure}


\bibliography{references.bib}